%% file: bootloops.tex
\documentclass[11pt]{article}
\usepackage[utf8]{inputenc}
\usepackage[T1]{fontenc}
\usepackage{lmodern}
\input{glyphtounicode}\pdfgentounicode=1
\usepackage[margin=1in]{geometry}
\usepackage{amsmath,amssymb,amsfonts}
\usepackage{xcolor}
\usepackage{graphicx}
\usepackage{calc}

\usepackage{booktabs}
\usepackage{array}
\usepackage{enumitem}
\usepackage[colorlinks=true,linkcolor=blue!55!black,citecolor=green!45!black,urlcolor=blue!55!black]{hyperref}
\usepackage{tikz}
\usetikzlibrary{positioning}
\usepackage{titlesec}
\usepackage{amsthm}
\input{fdiag-style.tex}

\newcommand{\eps}{\varepsilon}

\titleformat{\section}{\normalfont\Large\bfseries}{\thesection}{0.6em}{}
\setlist{itemsep=2pt,topsep=3pt}
\title{BootLoops: an LLM-driven toolkit for exact quantitative science}
\author{Matthew D. Schwartz\\Department of Physics, Harvard University}
\date{October 1, 2026}

\begin{document}
\maketitle

\input{abstract}

\input{s1_intro}
\input{s2_toolkit}
\input{s3_feynman}

\input{s4_integrals}
\input{s5_enumeration}
\input{s6_ball}
\input{s7_scope}

\section*{Acknowledgments}
We would like to thank the following people for advice and collaboration: Isaiah Andrews, Nima Arkani-Hamed, Michael Desai,
Ethan Dyer, Scott Edwards, Noam Elkies, Cecilia Garraffo, Matthew Gordon, Thomas Grimm,
Martin Hemberg, Mikhail Ivanov, David Johnston, Gary King, Paul Lewis, Amara McCune,
Brendan Meade, Siddharth Mishra-Sharma, James O'Dwyer, Joe Pater, Kevin Ryan,
Subhabrata Sen, Jesse Shapiro, and Xiaoyuan Zhang.

This paper was written with the assistance of generative AI, Claude Fable 5.1 (Anthropic). The BootLoops package was written by AI, except for forks of public packages which are their authors' work. The author takes full responsibility for the contents of this paper and of BootLoops.
Funding for this project was provided by Anthropic. This is not an official Anthropic project and results and views are not endorsed by Anthropic.

\bibliographystyle{unsrt}%
\bibliography{refs,refs_add}

\end{document}

%% file: fdiag-style.tex
\usetikzlibrary{decorations.pathmorphing,arrows.meta,calc}

\tikzset{
  fdiag/.style={
    line cap=round, line join=round,
    every node/.style={font=\small},
  },
  massless/.style={thin, black},
  masslessTHICK/.style={thick, black},
  massive/.style={ultra thick, red!70!black},      %
  massM1/.style={ultra thick, red!70!black},       %
  massM2/.style={ultra thick, blue!75!black},      %
  massM3/.style={ultra thick, green!55!black},     %
  massM4/.style={ultra thick, orange!90!black},    %
  massHeavy/.style={ultra thick, violet},          %
  ext/.style={thick, black, -{Latex[length=2mm]}},
  offshell/.style={thick, blue!60!black, decorate,
    decoration={snake, amplitude=0.6mm, segment length=2mm}, -{Latex[length=2mm]}},
  offshellH/.style={ultra thick, violet, -{Latex[length=2mm]}},
  lbl/.style={font=\footnotesize},
}

\newcommand{\vertexdot}[1]{\fill #1 circle (1.9pt);}

%% file: abstract.tex
\begin{abstract}
Much of quantitative science involves difficult mathematical computations: a Feynman integral, evidence for an evolutionary tree, or a regulatory statistic. These are often estimated by Monte Carlo or heuristic search, or computed in floating point without tracking rounding error. Exact methods often exist, but in a field far from the problem. The reduction methods of collider physics also work on a Bayesian evidence, and a regulatory threshold can be settled with interval arithmetic from numerical analysis, but few people know both fields. A large language model knows both, and every other field, and with working programs in hand it can carry a method from one to another. BootLoops is a pool of such programs: integral reduction, differential equations and high-precision evaluation from particle physics, integer-relation fitting from experimental mathematics, exact enumeration, and ball arithmetic, in one toolkit that any agentic large language model can operate and extend. These tools can compute multi-loop Feynman integrals by bootstrap methods rather than direct integration, across the function classes that arise, including polylogarithms, elliptic functions, and periods of K3 surfaces and Calabi--Yau manifolds. They can also evaluate Bayesian evidence integrals as exact rational numbers, enumerate finite configuration spaces with proof of completeness, show when a sought closed form cannot exist, and redo floating-point calculations with guaranteed digits. In practice this brings the tools of mathematics, computer science and physics to problems in genomics, statistics, ecology and other fields. The toolkit and its documentation are at \url{https://bootloops.ai}.
\end{abstract}

%% file: s1_intro.tex
\section{Introduction \label{sec:intro}}

For most of the history of scientific computing, a human being would write a program, knowing exactly what it does, and then a computer would carry it out. With machine learning, this paradigm changed: a human would specify the architecture, a computer would fit a neural network, and the human would interpret the results. For example, AlphaFold predicts the structure of a protein from its sequence~\cite{AlphaFold2021}, GraphCast produces medium-range weather forecasts~\cite{GraphCast2023}, and networks trained on known crystals predict new stable materials~\cite{GNoME2023}. These neural networks can stand in for a calculation that is too expensive to run or that nobody knows how to write down otherwise. Now the paradigm has changed again: a large language model running as a semi-autonomous artificially intelligent agent can design and write the program, run it, and review, rewrite and interpret the answer itself. The potential of such agentic systems seems unbounded, yet the mechanism by which humans can be maximally productive using agentic AI remains uncertain. There is, however, one clear path by which the utility of agentic AI can be enhanced: allowing the models to bring together techniques from distant fields. In particular, methods developed for computer science, mathematics and physics can be directly relevant for problems in population genetics or ecology, yet few people know both sides well enough to connect them. This paper introduces BootLoops, a harness for large language models to use that brings together various packages for scientific computing in one place.

One of the great strengths of large language models is the breadth of their knowledge. In science, it may be possible to compute a quantity of interest exactly, or with a proven bound on its error, by a method that already exists in a field far removed from the problem, which no human has noticed but an AI can find. For example, the sophisticated reduction methods that physicists and mathematicians developed for Feynman integrals apply directly to Bayesian evidence integrals needed by statisticians and biologists. Or an interest group may want to know if some published data clears a regulatory threshold after being processed through a mandated sequence of algorithmic steps, but the rounding error grows uncontrollably. With interval arithmetic, a tool from numerical analysis that carries a proven error bound through every step, a rigorous reading may be possible. Making these kinds of connections requires understanding the mathematics or computer science from one field and the questions of another, connections that may be hard for humans to make but easy for large language models.

Even if language models are capable of making the connections, they still need to implement and port the methods from one field to another. This takes work. But the work does not have to be repeated over and over again by every scientist interested in the connection. The imported and novel code can be centralized and made accessible. BootLoops serves this purpose: it brings programs for quantitative science together from distant fields so that a language model can run them. These programs reduce and evaluate integrals, solve their differential equations, fit exact constants to high-precision numbers by integer-relation methods, enumerate finite sets exhaustively, and compute in exact and in ball (interval) arithmetic. Some of them are public codes, which we adopted as released or, where they were written as Mathematica packages, translated to run on our own open numerical libraries (Section~\ref{sec:growth}). More of the work went into improving existing codes, which the model, Claude (Anthropic), extended and in places rebuilt when their numerical methods failed at the precision required. Most of the programs, however, are new, and many of them implement methods described in papers but never released as code, for example in the theory of modular forms and in algebraic statistics. The others the model wrote from scratch, where no suitable method existed. For the Feynman-integral computations alone, roughly one hundred new programs were written.

The initial version of BootLoops has tools for many computations in four broad categories. First, it has tools to evaluate integrals, such as for Feynman diagrams of scattering amplitudes, in closed form by bootstrap methods, without direct integration. Past one loop the answers are in general no longer polylogarithms, and the functions that appear are governed by elliptic curves, K3 surfaces and Calabi--Yau manifolds~\cite{SAGEXch3}. In the S-matrix bootstrap one writes down the most general function in an allowed space and lets the known singularities and limits fix the free coefficients~\cite{HexagonBootstrap}. BootLoops combines the Landau bootstrap, which uses the singularity locus of a single integral~\cite{LandauBootstrap}, with analytic regression, which fits a formula to high-precision values by integer-relation methods~\cite{AnalyticRegression}, for a powerful semi-numerical bootstrap apparatus. An advantage of the semi-numerical approach is that each closed form can be checked against an independent numerical evaluation of its integral to hundreds of digits if needed. The same machinery developed for scattering amplitudes also works for period integrals in cosmology, gravitation and string theory. Second, BootLoops can compute Bayesian evidence integrals for count data under a polynomial model. These integrals are the same mathematical objects as Feynman integrals in their Lee--Pomeransky representation, and BootLoops can evaluate them as exact rational numbers for many statistical models. Third, BootLoops can enumerate finite configuration spaces with a rigorous proof that no case was missed. Where a closed form does not exist, BootLoops can prove it. Finally, BootLoops can redo general floating-point calculations in ball arithmetic, so that error propagation is under control and exact bounds can be established.

BootLoops has many applications. Phylogeneticists compare candidate trees by their Bayesian evidence~\cite{PhyloNote}, as do statisticians for mixture models~\cite{MixturesNote}. Related likelihoods arise in population genetics, in ecology and in models of transcription in single cells~\cite{GeneburstingNote}. String theorists counting flux vacua~\cite{StringvacNote} face a finite set of cases that can be enumerated completely, as do the designers of an adaptive clinical trial~\cite{HealthtrialsNote}.
Exact and ball arithmetic can also be used when numbers follow from a stated rule, as in Medicare's star ratings~\cite{HealthstarsNote} and the U.S. ozone standard~\cite{AirqualityNote}.

Building a harness to make agentic AI more effective is not a new idea. For example, FunSearch pairs a language model with a program that scores the model's proposals, leading to new constructions in extremal combinatorics~\cite{FunSearch2024}. The AI Scientist runs a language model as an agent through a machine-learning project, from the idea to the written paper~\cite{AIScientist2024}. More recently, Robin, a system of several agents built at FutureHouse, ran the cycle of literature search, hypothesis, experiment design and data analysis for a problem in biology and proposed a drug candidate for dry macular degeneration that laboratory experiments then supported~\cite{Robin2026}. These systems, however, are aimed at autonomous science. BootLoops is aimed at a language model working under human supervision. It is a collection of tools that people would use if they knew of them, and the language model makes them easy to use even for someone who does not fully understand how they work. In this it resembles Claude Science, a harness that Anthropic built to make agentic AI easier for scientists to use~\cite{ClaudeScience2026}. Importantly, though, BootLoops can be used with any underlying language model, or inside a commercial harness such as Claude Science or Codex (OpenAI)~\cite{Codex2025}, and it is open source and can be extended by the community of scientists who use it. 

No single ingredient of the methods in BootLoops is new. The reduction algorithms and high-precision numerics come from collider physics, the machinery of periods from mirror symmetry, the modular-form and integer-relation methods from number theory and experimental mathematics, and the evidence integrals from statistics. Interval arithmetic and exhaustive search with proof are standard tools of numerical analysis and computer science. Our contribution is to pool these methods as working programs in a single toolkit, BootLoops, which a language model can operate and extend because it knows all of these fields. Section~\ref{sec:toolkit} describes the toolkit and its growth, Sections~\ref{sec:amplitudes}--\ref{sec:ball} the four kinds of computation in turn, and Section~\ref{sec:outlook} concludes.

%% file: s2_toolkit.tex
\section{The toolkit \label{sec:toolkit}}

A computation with BootLoops involves a large language model and the toolkit of deterministic programs that the model operates and extends. The programs implement methods of particle physics, experimental mathematics, computer algebra and numerical analysis (Table~\ref{tab:toolkit}), and we use them for the four kinds of computation listed in Table~\ref{tab:types}. This section describes the toolkit and contains no new results.

\subsection{The model and the toolkit \label{sec:system}}

The computations reported here were run by a large language model, Claude (Anthropic), with a human setting the targets and the standards. The model runs as an agent, that is, in a loop in which it issues commands that execute programs and then reads their output. Given an object to compute, the model plans the calculation and runs the programs that each step requires. On a new integral family, for example, it classifies the geometry (sphere, elliptic curve, K3 or Calabi--Yau), proposes the space of functions in which the answer should lie, and runs the reductions, transports and fits that fix the answer there. When the existing tools lack an operation that a problem requires, the model writes a new routine, tests it on cases with known answers and adds it to the toolkit. The programs compute every number and symbol. At no step is the model asked for a digit, and no result rests on the model's judgment.

The toolkit persists across problems. It holds the programs that reduce an integral family to a finite basis, solve its differential equations, evaluate it numerically, fit function spaces to the values, integrate with proven error bounds and enumerate configurations exhaustively. The programs share formats, and the output of one becomes the input of the next. None of this changes the model itself: the weights never move. The improvement from one problem to the next accumulates in the toolkit and its documentation. A harness, as the term is used for language models, is the layer of software through which a model is operated. BootLoops is a living harness in this sense, since people and the model extend it with every problem while it stays separate from the model.

\subsection{The toolkit by class and its growth \label{sec:growth}}

Table~\ref{tab:toolkit} lists the principal programs of the toolkit in nine classes by function. The toolkit's external part consists of public programs. Some, such as the reduction program Kira~\cite{KiraIBP} and the Landau-singularity program PLD~\cite{PLDjl}, are used as released (Kira through a performance-patched fork whose reduction output is unchanged). Others, such as the singular-locus program SOFIA~\cite{SOFIA} and the reduction program Blade~\cite{Blade2025}, were published as Wolfram Language packages. Such a package is its authors' own code: it implements the physics and calls Mathematica's built-in routines for generic tasks such as arbitrary-precision arithmetic, linear algebra, the series solution of differential equations, polynomial algebra and numerical integration. To port such a package we wrote our own libraries for those generic tasks, in C++ and Julia, from the standard published algorithms, and translated the authors' package code to call them. A port therefore consists of the package authors' algorithms running on our open libraries; SOFIA runs this way in Julia and Blade on a compiled toolchain. For auxiliary-mass flow~\cite{AMFlow} a public C++ reimplementation, AMFlow.cpp~\cite{AMFlowCpp}, already existed, and we adopted it as a fork that we extended. No required step then needs a Mathematica license, and a language model can read, run and modify every layer of the program. No attempt was ever made to reverse-engineer any routine in a commercial package; code for ports was written from scratch using public algorithms.

Where we changed a public open-source program's numerical methods or added routines, we keep the changed version as a fork (a full copy of the original project carrying our patches). The original code, license and attribution then stay intact, and the patches can be offered back. The internal part is the larger one, and we wrote it in the course of the work. For the Feynman-integral computations alone this amounted to roughly one hundred purpose-built tools where no public equivalent existed.

\begin{table}[tp]
\centering
\setlength{\abovecaptionskip}{0pt}\setlength{\belowcaptionskip}{5pt}
\caption{The toolkit by function: in each class the mature public programs
were adopted as released or ported, and the remaining programs were
written in the course of the work where nothing public covered the need.
Under each class name we give the sections where the class is used
and, for a program reused in another kind of
computation, the section where it is reused. Names are the programs' public names; every program has a
documentation page in the online registry at \url{https://bootloops.ai} (Section~\ref{sec:outlook}).
Programs of the last class check the output of the other eight.}
\label{tab:toolkit}
\scriptsize
\setlength{\tabcolsep}{3pt}
\begin{tabular}{@{}>{\raggedright}p{2.6cm} >{\raggedright}p{5.3cm} >{\raggedright}p{3.3cm} >{\raggedright\arraybackslash}p{4.6cm}@{}}
\toprule
Class (used in) & What it computes & Adopted or ported & Written for BootLoops \\
\midrule
Reduction engines (Sec.~\ref{sec:amplitudes}) &
reduction of the thousands, sometimes millions, of integrals in a family
to a finite basis by integration-by-parts identities; the memory a
reduction will need, predicted before it runs &
Kira~\cite{KiraIBP} with FireFly~\cite{FireFly2020} and Fermat (a performance-patched fork, reduction output unchanged); Blade~\cite{Blade2025}, ported
to run without Mathematica &
Seedling~\cite{SeedlingNote} (staged seeding, cost prediction); Winnow (each identity with its own proof; one sector's differential equation from a family too large to reduce whole) \tabularnewline\addlinespace
Singularities and geometry (Sec.~\ref{sec:amplitudes}) &
the complete set of points where an integral can be singular, the symbol
alphabet it implies, and the geometry of the maximal cut (sphere, elliptic
curve, K3, Calabi--Yau) that fixes the class of functions &
PLD~\cite{PLDjl}, as released; SOFIA~\cite{SOFIA}, ported to Julia &
Leviathan (independent second computation of the singular set); Landau
Alphabet (singular set to alphabet); GeoTriage (geometry from the
Picard--Fuchs operator); Maxcut \tabularnewline\addlinespace
Differential equations and transport (Sec.~\ref{sec:amplitudes};
Annihilator reused in Secs.~\ref{sec:evidence}, \ref{sec:enumeration}) &
canonical form of a differential-equation system; the connection
reconstructed from numerical samples; a value carried from one kinematic
point to any other along the equation; the annihilating differential or
recurrence operator of a series, with minimal order proven &
none &
Counterweight; Wayfinder; Vopclose; Annihilator \tabularnewline\addlinespace
Evaluators (Sec.~\ref{sec:amplitudes}) &
master integrals to hundreds of digits; polylogarithms and elliptic, K3
and Calabi--Yau periods to any requested precision; quadrature with a
proven error bound &
AMFlow.cpp~\cite{AMFlowCpp}, the public C++ reimplementation of auxiliary-mass flow~\cite{AMFlow}, forked and extended; pySecDec~\cite{pySecDec2018}
and FIESTA~\cite{FIESTA5}, as released, kept because they share no code with the rest &
Nestor (quadrature); GPLEval (polylogarithms); Eichler (periods)
\tabularnewline\addlinespace
Closure and naming (Sec.~\ref{sec:amplitudes}) &
the exact combination of declared constants that a high-precision value
equals, under a stated bound on the integer coefficients, or the statement
that none exists below the bound &
PSLQ~\cite{PSLQ1999} and LLL~\cite{LLL1982} &
Ansatzer (the ansatz and its residual count); Lockpick (lattice reduction
where PSLQ stalls); RingMaker (the admissible ring of constants);
Frobenius (boundary constants by arithmetic modulo primes)
\tabularnewline\addlinespace
Ball arithmetic (Secs.~\ref{sec:evidence}, \ref{sec:ball}) &
every real quantity as a midpoint with a proven radius; quadrature,
transport and root isolation with proven bounds; usable bounds when one
parameter enters every factor of a large product &
Arb~\cite{Johansson2017} and mpmath~\cite{mpmath} &
Baller~\cite{BallerNote}; ERAS \tabularnewline\addlinespace
Evidence and statistics (Sec.~\ref{sec:evidence}) &
the Bayesian evidence of a tree or a mixture model as an
exact number or a proven enclosure; population-genetic likelihoods with
proven error bounds &
none &
JaCK \& Jill~\cite{JackandjillNote}; Mixalot~\cite{MixturesNote}; Popcorn
\tabularnewline\addlinespace
Enumeration and audits (Secs.~\ref{sec:enumeration}, \ref{sec:ball}) &
a complete enumeration with a proof that nothing was missed; the exact
replay of published arithmetic from the publication's own inputs &
none &
Terrier~\cite{StringvacNote} (flux lattices, exact lattice arithmetic);
Recount (replay of published numbers; distributed with the audits) \tabularnewline\addlinespace
Integrity (all sections) &
the precision and provenance of every reference value; reduction output
confirmed by checks with disjoint code; an integral family confirmed
against its graph before any reduction runs &
none &
Gatekeeper; Trust; the family preflight of Seedling \tabularnewline
\bottomrule
\end{tabular}
\end{table}

\begin{table}[tp]
\centering
\setlength{\abovecaptionskip}{0pt}\setlength{\belowcaptionskip}{5pt}
\caption{The four kinds of computation of
Sections~\ref{sec:amplitudes}--\ref{sec:ball}, all carried out with the
programs of Table~\ref{tab:toolkit}: the object computed exactly in each,
the usual practice in the fields concerned, and the areas of application.
Results are stated in this paper for the first row and, in
Section~\ref{sec:impossible}, for the Watson integral of the third; otherwise
Sections~\ref{sec:evidence}--\ref{sec:ball} describe the computation and its
relevance to the areas named without reporting results.}
\label{tab:types}
\footnotesize
\setlength{\tabcolsep}{3pt}
\begin{tabular}{@{}>{\raggedright}p{3.0cm} >{\raggedright}p{4.4cm} >{\raggedright}p{3.8cm} >{\raggedright\arraybackslash}p{4.4cm}@{}}
\toprule
Kind of computation & Exact object & Usual practice in the fields & Application areas \\
\midrule
Function-space bootstrap of Feynman-class integrals
(Sec.~\ref{sec:amplitudes}) &
a closed form $F(s)$ for an integral of the class of
Eq.~\eqref{eq:eulerclass}: named functions on a stated alphabet, with
every constant classical or given by a convergent integral, and a
standalone evaluator &
numerical evaluation point by point~\cite{AMFlow}; closed forms derived
case by case &
collider amplitudes; energy correlators; gravitational-wave scattering;
cosmological correlators; large-scale structure \tabularnewline\addlinespace
Exact evidence and the holonomic class (Sec.~\ref{sec:evidence}) &
the evidence $Z$ of Eq.~\eqref{eq:evidence} as an exact rational number
or an element of a stated ring; more generally a period or holonomic
function together with its annihilating operator &
Monte Carlo estimates of the evidence~\cite{LartillotPhilippe2006,Xie2011,Fourment2020},
whose error bars are themselves estimates &
phylogenetics; mixture models and authorship; forest dynamics;
single-cell transcription kinetics; population genetics
\tabularnewline\addlinespace
Exhaustive enumeration and proof (Sec.~\ref{sec:enumeration}) &
the complete list of configurations a theory allows, or a theorem that no
object of a stated form exists (an operator of proven minimal order) &
sampled or heuristic search; conjectured closed forms tested numerically &
string flux vacua; Voynich readings; the Watson integral;
clinical-trial operating characteristics; clustering optima of rating
rules \tabularnewline\addlinespace
Ball arithmetic and audits (Sec.~\ref{sec:ball}) &
a real number as a ball $m\pm r$ of Eq.~\eqref{eq:ball} whose radius is
a proven bound; for a published number, its recomputation from the
publication's own inputs &
floating-point arithmetic whose rounding error is not tracked; reruns at
higher precision by hand &
Medicare star ratings; the ozone standard's truncation arithmetic;
population-genetics reference software; periods of string
compactifications \tabularnewline
\bottomrule
\end{tabular}
\end{table}

New tools were added to the toolkit in four ways. A public program was ported in the sense defined above, as with SOFIA and Blade. A program reached a limit in the middle of a problem and was upgraded (the upgrade kept as a fork). For example, we rebuilt the series solver of the AMFlow.cpp fork to keep nearly coincident solution families separated at very high working precision. The model turned methods that existed only in papers into code: Eichler evaluates the Eichler integrals (iterated integrals of modular forms~\cite{Eichler1957}) that appear in elliptic Feynman integrals, and Mixalot computes the evidence of a mixture model exactly by methods from algebraic statistics~\cite{LinSturmfelsXu2009}.

Finally, a tool was built from scratch where nothing existed. Public programs find where an integral is singular~\cite{PLDjl,SOFIA}, but none that we found turned that singular set into the working ansatz of the bootstrap (the alphabet, the analytic conditions on it, and the count of coefficients left to fit). We wrote the pair Landau Alphabet and Ansatzer for this step. The interval routine ERAS (Section~\ref{sec:eras}) was written for the same reason during a statistical application, when plain interval arithmetic returned unusably wide bounds on a likelihood.

A tool built for one problem has usually been reused in later ones, either unchanged or extended. Eichler was written to evaluate one function, was then extended to the periods of a family of elliptic curves, and later to K3 and Calabi--Yau periods. Annihilator, written to find and certify the differential and recurrence operators of Feynman-integral series, computes the recurrences of evidence integrals unchanged. The routines and documentation written during one problem remain in the toolkit for the next. Because the toolkit is separate from the model, any agentic large language model, including a later one, should be able to operate and extend it as it stands.

\subsection{Documentation and instructions \label{sec:protocol}}

Every program in the toolkit has a documentation page, and the toolkit is distributed with plain instruction files that any model can read. These state what each tool does, when to reach for it, what its output means and how its answer is tested. The intended use is that a large language model will read these files first. If the tools exist that it needs for its project, great. If not, it locates where an existing tool applies and extends that tool instead of writing another from scratch if possible. Working protocols for good scientific practice and agentic AI use are distributed with BootLoops as well. For example, one protocol says to try a method on planted answers before real data. Another recommends recording the source and precision of every reference value.

Being able to check and validate results in a semi-autonomous way is critical to productive use of agentic AI. The validation required naturally varies with the kind of computation. One particularly clean example described at length in this paper is the check on closed forms for integrals (Section~\ref{sec:method-solved}). There one may use numerical evaluation to help find the analytic answer, but the result then must be validated to many digits of precision beyond what was used in the computation.

%% file: s3_feynman.tex
\section{Feynman integrals and amplitudes\label{sec:amplitudes}}

Loop integrals in quantum field theory can be computed in closed form without direct integration, from high-precision numerical values and constraints that the answer must satisfy. Fixing an answer from general principles alone is an old idea in particle physics. In the 1950s and 60s physicists tried to determine scattering amplitudes from analyticity and unitarity without a Lagrangian, an approach that Geoffrey Chew called the bootstrap~\cite{Eden:1966dnq}. In the perturbative version used here we determine one integral at a time. We describe the class of integrals, the bootstrap and its acceptance test, and the thirty integrals computed so far.

\subsection{The problem class\label{sec:method-structure}}

A scalar Feynman integral with $L$ loops and $N$ internal lines is, in momentum space,
\begin{equation}
I(a_1,\ldots,a_N)\;=\;\int \prod_{\ell=1}^{L}\frac{d^d k_\ell}{i\pi^{d/2}}\;\frac{1}{D_1^{a_1}\,D_2^{a_2}\cdots D_N^{a_N}},\qquad D_j=q_j^2-m_j^2
\label{eq:feynman}
\end{equation}
with one integration per loop momentum $k_\ell$. Each internal line contributes a propagator $D_j$, the factor for a virtual particle of mass $m_j$ and momentum $q_j$ (a fixed combination of loop and external momenta), raised to an integer power $a_j$. In $d=4-2\eps$ dimensions divergences become poles in $\eps$, and each order in $\eps$ is a function of the kinematic invariants $s$ (masses and external momenta).

Feynman parametrization brings every such integral to the form
\begin{equation}
I(s;\eps) \;=\; \int_\Delta d^n x \;\prod_{i} P_i(x;s)^{\,\nu_i + a_i \eps}\,,
\label{eq:eulerclass}
\end{equation}
a generalized Euler integral~\cite{GKZ1990}. Here $\Delta$ is a semialgebraic domain in the $n$ variables $x$, the $P_i$ are polynomials, and the exponents $\nu_i+a_i\eps$ are fixed for each problem and linear in $\eps$. For a Feynman graph the $P_i$ are the two Symanzik polynomials $\mathcal U$ and $\mathcal F$, built from the spanning trees and two-forests of the graph with the invariants $s$ as coefficients. Their exponents follow from the propagator powers $a_j$ of Eq.~\eqref{eq:feynman}, the number of loops $L$ and the dimension $d$. In the Lee--Pomeransky representation the single polynomial $\mathcal G=\mathcal U+\mathcal F$ suffices~\cite{LeePomeransky}.

The possible singularities of $I(s;\eps)$ in $s$ are known in advance. They lie on algebraic varieties, the Landau loci, where the zero sets of the $P_i$ can pinch the integration domain, and are computable from the integrand alone~\cite{LandauBootstrap,MizeraTelenLandau}. The class~\eqref{eq:eulerclass} is also holonomic: integration by parts (IBP) yields linear relations among integrals whose exponents differ by integers, and these relations form a closed system. In particular, every family (fixed propagators $D_j$, all integer powers $a_j$) is spanned by finitely many \emph{master integrals}, which satisfy a first-order linear system of differential equations in $s$ with rational coefficients~\cite{Laporta:2001dd,KiraIBP}.

The function space of such an integral is fixed by the algebraic variety that the zero sets of the $P_i$ define. The maximal cut of a master integral (its residue on the poles of all propagators at once) is a period of this variety. Its Picard--Fuchs operator (the differential operator in $s$ annihilating the period) identifies the type of the variety. A rational variety gives multiple polylogarithms~\cite{Goncharov:2010jf,HennCanonical}; a modular elliptic curve, iterated integrals of modular forms~\cite{AdamsWeinzierlSunrise,BlochVanhove2014}. The K3 surfaces and Calabi--Yau varieties next in dimension give the periods computed in mirror symmetry~\cite{Candelas:1990rm,BananaAllLoop}, and the boundary constants of the differential equations are periods too~\cite{KontsevichZagier}.

Closed forms for these integrals are hard to produce and easy to check. A Feynman integral in the class can usually be evaluated to any needed precision by auxiliary-mass flow~\cite{AMFlow}, independently of any closed form. This method gives the propagators an auxiliary mass and transports the integral numerically through its differential equation from large mass to zero.

\subsection{The function-space bootstrap\label{sec:method-pipeline}}

\begin{figure}[t]
\centering
\input{fig-system.tex}
\caption{The bootstrap of one Feynman integral with the toolkit.
(a)~The parts of the computation. The model operates and extends the
toolkit, which persists across problems and is separate from the model.
The programs produce a candidate closed form $F(s)$, which is compared
with an independent numerical evaluation $I^{\rm num}(s_k)$
(auxiliary-mass flow~\cite{AMFlow} on a separate reduction~\cite{KiraIBP},
used in no fit) at points $s_k$ that entered no stage of the derivation,
Eq.~\eqref{eq:accept}. (b)~The squeeze on one integral. The geometry of
the singularities bounds the space of candidate functions from above, and
high-precision samples with integer-relation fits fix the remaining
coefficients from below.}
\label{fig:squeeze}
\end{figure}
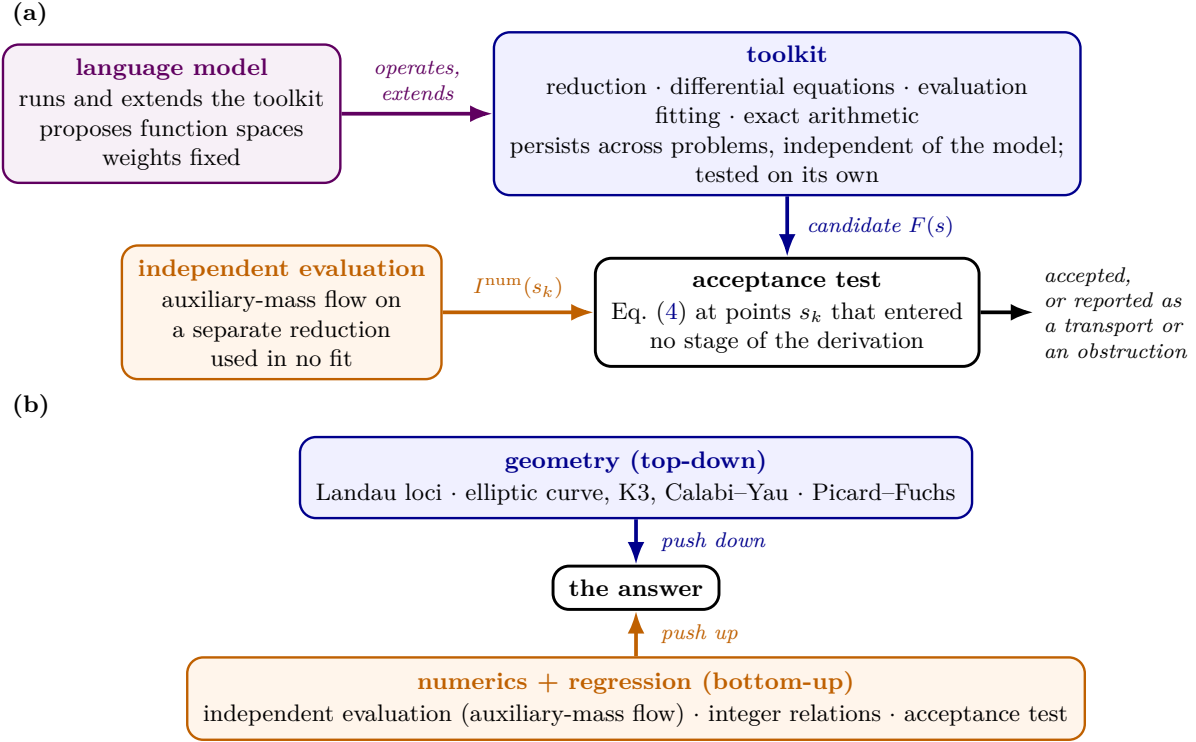

On a single integral the bootstrap (Fig.~\ref{fig:squeeze}a) is the two-sided squeeze of Fig.~\ref{fig:squeeze}b, which we run in five stages. First, we compute the complete Landau singular locus from the graph alone~\cite{PLDjl,SOFIA}. Its components give the alphabet of the answer, a finite list of letters (algebraic functions of $s$). Second, we classify the maximal cut of each master integral as genus zero (a sphere), elliptic, K3 or Calabi--Yau, which fixes the iterated-integral kernels. With the alphabet and the leading singularity (the algebraic prefactor), the kernels generate a weight-graded ansatz: every allowed function of the right transcendental weight (one for a logarithm, two for a dilogarithm) with an unknown coefficient. The routines for these stages are Landau Alphabet and Ansatzer. Third, exact constraints (branch points only on the Landau loci, first-entry conditions, integrability, the symmetries of the graph) reduce the ansatz to $10^2$--$10^3$ unknown coefficients. Indeed, for polylogarithmic master integrals these constraints often fix the answer outright.

Fourth, we constrain the surviving unknowns with high-precision samples of the integral, from auxiliary-mass flow~\cite{AMFlow} on an IBP reduction~\cite{KiraIBP} or, beyond genus zero, from transport of the periods with rigorous error control. Fifth, we recognize the constants by integer-relation fits. Given real numbers $x_1,\dots,x_n$ known to $D$ digits, the PSLQ and LLL algorithms~\cite{PSLQ1999,LLL1982} find integers $a_i$, not all zero, with
\begin{equation}
a_1x_1+a_2x_2+\cdots+a_nx_n=0
\label{eq:intrel}
\end{equation}
to the working precision, or establish that no relation exists with every $|a_i|$ below a stated height $H$. We accept a relation only if its height is far below the largest that $D$ digits can resolve, it is stable at two working precisions, and a planted control is recovered. Where the acceptance test below cannot be passed, we report what was established instead: a value carried along the differential equation with a proven error bound, a closed form with some constants left as their defining integrals, or a proof that no function of the proposed class equals the integral.

The five stages form the \emph{constraint leg} of the method. The other leg is the \emph{integration leg}: reduce to master integrals, then solve their differential equations from an exactly known boundary value. That is, we can also compute the integral by solving its differential equations. Sometimes this is the easiest route, and BootLoops has many packages built and ported for this route as well as for the bootstrap. We assign each master integral to a leg before any sampling, chiefly by whether its values can in principle be fit. An elliptic curve has two independent periods, and the maximal cut supplies only the first. Thus if the normalization of a master integral shows it mixes the two periods, a fit with constant coefficients provably cannot match it at all kinematic points, and that master is assigned to the integration leg.

Three further routines serve both legs: rotation of the differential equations to an $\eps$-factorized basis where one exists~\cite{HennCanonical,LeeEpsForm}, reconstruction of exact rational functions from finite-field evaluations~\cite{Peraro2016}, and representation of an integral with no closed form by its annihilating differential or difference operator, found modulo primes, lifted to the rationals and proved by creative telescoping~\cite{Koutschan2013}.

\subsection{The acceptance test\label{sec:method-solved}}

Let $F(s)$ be a candidate closed form for an integral of the type~\eqref{eq:eulerclass}. Let $I^{\rm num}(s)$ be an evaluation of the same integral obtained independently of the derivation of $F$, in most cases auxiliary-mass flow~\cite{AMFlow} on a separate reduction~\cite{KiraIBP}. We accept $F$ when
\begin{equation}
\frac{\bigl|F(s_k)-I^{\rm num}(s_k)\bigr|}{\bigl|I^{\rm num}(s_k)\bigr|}\;<\;10^{-30}\qquad\text{for every held-out point } s_k\,,
\label{eq:accept}
\end{equation}
where the $s_k$ are kinematic points that entered no stage of the derivation. Agreement of two independently computed real numbers to thirty digits is a coincidence with probability $\sim 10^{-30}$, and each acceptance amounts to a hypothesis test at that significance. Thirty digits is the floor, and many results are tested at sixty or a hundred. Keep in mind that the evidence for an accepted closed form is numerical, in the tradition of high-precision identifications~\cite{Laporta:2001dd,Laporta4loop}; thus integrals computed this way do not constitute a symbolic proof.

A numerical value of an integral, however precise, does not show which functions its closed form is built from. The {\bf BootLoops standard} for Feynman integrals requires three things of a computed integral: an explicit symbolic form over a declared alphabet, with every constant classical or defined by a written-out convergent integral; a standalone evaluator of that form at any kinematic point and precision from arbitrary-precision arithmetic alone; and Eq.~\eqref{eq:accept}, passed at two working precisions. Auxiliary-mass flow supplies the samples of the fourth stage and, from a separate reduction, the independent side of Eq.~\eqref{eq:accept}; it also serves on some transport routes, and it never appears inside a closed form itself.

We search for a classical name for every boundary constant, but one need not exist. The period of a generic elliptic curve, one without complex multiplication whose family over the kinematic line is not modular, is not expected to lie in any classical ring. When no relation is found below a stated height, the constant enters the closed form through its defining integral~\cite{SmatrixPaper}.

\subsection{Thirty integrals in closed form\label{sec:smatrix}}

With BootLoops we have computed thirty Feynman integrals, from one to four loops, in symbolic closed form~\cite{SmatrixPaper,FeynmanbootstrapNote,FeynmangeometryNote}, ranging over polylogarithmic, elliptic, K3 and Calabi--Yau geometries. All thirty (Table~\ref{tab:portfolio}) meet the standard of Section~\ref{sec:method-solved}. Fifteen of the thirty were already computed in the literature, among them the four-loop equal-mass banana.
For these fifteen the published closed forms were used only at the final comparison, where all fifteen agree; what each computation took from the literature, if anything, is stated in its entry in Ref.~\cite{SmatrixPaper}. The other fifteen are new: we have found no prior computation of them in the literature~\cite{SmatrixPaper}. Three of the known and twelve of the new integrals are drawn in Fig.~\ref{fig:gallery15}. Twenty-four of the thirty were fixed by geometry, analytic constraints and integer-relation fits, and six went through the integration leg. Table~\ref{tab:portfolio} lists each integral's agreement with held-out values (one qualification appears in its caption), and Ref.~\cite{SmatrixPaper} contains one derivation per graph.

\input{tab-portfolio}

\input{gallery15}

%% file: fig-system.tex
\begin{tikzpicture}[
  every node/.style={font=\footnotesize},
  box/.style={rounded corners=6pt, line width=0.9pt, align=center,
              inner xsep=6pt, inner ysep=5pt, draw=black},
  flow/.style={line width=1.4pt, -{Latex[length=3mm]}},
  lab/.style={font=\scriptsize\itshape, align=center},
  node distance=8mm and 2cm]
  \node[box, draw=blue!55!black, fill=blue!6] (kit)
    {\textbf{\textcolor{blue!55!black}{toolkit}}\\[1pt]
     reduction $\cdot$ differential equations $\cdot$ evaluation\\
     fitting $\cdot$ exact arithmetic\\
     persists across problems, independent of the model;\\ tested on its own};
  \node[box, draw=violet!75!black, fill=violet!6, left=of kit] (model)
    {\textbf{\textcolor{violet!75!black}{language model}}\\[1pt]
     runs and extends the toolkit\\ proposes function spaces\\
     weights fixed};
  \node[box, line width=1.1pt, below=of kit] (test)
    {\textbf{acceptance test}\\[1pt]
     Eq.~\eqref{eq:accept} at points $s_k$ that entered\\
     no stage of the derivation};
  \node[box, draw=orange!75!black, fill=orange!8, left=of test] (indep)
    {\textbf{\textcolor{orange!75!black}{independent evaluation}}\\[1pt]
     auxiliary-mass flow on\\ a separate reduction\\ used in no fit};
  \node[lab, align=left, right=7mm of test] (out)
    {accepted,\\ or reported as\\ a transport or\\ an obstruction};
  \draw[flow, violet!75!black] (model.east) -- (kit.west)
    node[midway, above=1pt, lab, violet!75!black] {operates,\\ extends};
  \draw[flow, blue!55!black] (kit.south) -- (test.north)
    node[midway, right=3pt, lab, blue!55!black] {candidate $F(s)$};
  \draw[flow, orange!75!black] (indep.east) -- (test.west)
    node[midway, above=1pt, lab, orange!75!black] {$I^{\rm num}(s_k)$};
  \draw[flow] (test.east) -- (out.west);
  \node[font=\footnotesize\bfseries, anchor=south west]
    at ($(model.north west)+(0,1mm)$) {(a)};
  \node[box, draw=blue!55!black, fill=blue!6, below=9mm of test,
        xshift=-2cm] (top)
    {\textbf{\textcolor{blue!55!black}{geometry (top-down)}}\\[1pt]
     Landau loci $\cdot$ elliptic curve, K3, Calabi--Yau $\cdot$ Picard--Fuchs};
  \node[box, line width=1.1pt, below=6mm of top] (ans)
    {\textbf{the answer}};
  \node[box, draw=orange!75!black, fill=orange!8, below=6mm of ans] (bot)
    {\textbf{\textcolor{orange!75!black}{numerics + regression (bottom-up)}}\\[1pt]
     independent evaluation (auxiliary-mass flow) $\cdot$ integer relations
     $\cdot$ acceptance test};
  \draw[flow, blue!55!black]   (top.south) -- (ans.north)
    node[midway, right=5pt, lab, blue!55!black] {push down};
  \draw[flow, orange!75!black] (bot.north) -- (ans.south)
    node[midway, right=5pt, lab, orange!75!black] {push up};
  \node[font=\footnotesize\bfseries, anchor=south west]
    at ($(model.north west |- top.north)+(0,1mm)$) {(b)};
\end{tikzpicture}

%% file: tab-portfolio.tex
\begin{table}[p]
\centering
\setlength{\abovecaptionskip}{0pt}\setlength{\belowcaptionskip}{5pt}
\caption{The thirty Feynman integrals computed in closed form, by the geometry of the maximal cut, numbered as in Ref.~\cite{SmatrixPaper}.
Fifteen were known and fifteen are new, with no prior computation that we could find (three and twelve of these appear in Fig.~\ref{fig:gallery15}). The order-$\eps^1$ sunrise and ice-cream cone count as rows.
Digits: agreement with comparison values that entered no fit (the floor is thirty; most elliptic rows reach 100 or more).
One qualification: for Nos.~2 and~4 the digits quoted are agreement with the analytic results of the literature, the auxiliary-mass-flow comparisons of those two rows reaching $\geq 89$ and $72$ digits.
$^{a}$Rational coefficients of the published form reproduced exactly.
$^{b}$Precision to which the closed-form identification is verified in ball arithmetic~\cite{FeynmangeometryNote}; the independent check of these two rows, against the IBP operator (Ref.~\cite{SmatrixPaper}), reaches 47 and 45 digits.}
\label{tab:portfolio}
\scriptsize
\renewcommand{\arraystretch}{1.02}
\begin{tabular}{@{}r >{\raggedright}p{3.7cm} >{\raggedright}p{8.6cm} >{\raggedright\arraybackslash}p{1.7cm}@{}}
\toprule
 & Integral & Description & Digits \\
\midrule
\multicolumn{4}{@{}l}{\emph{Polylogarithmic (maximal cut of genus zero)}}\\
1 & outer-mass double box & planar two-loop double box, one common mass on the six perimeter lines, massless legs & $\geq 86$ \\
2 & massless double box & planar two-loop double box, all lines massless & exact$^{a}$ \\
4 & three-loop ladder & planar massless ladder at $L=3$ & $\geq 461$ \\
5 & one-loop pentagon & massless, five on-shell legs; 15 letters already at weight two & 68 \\
6 & Sudakov vertex & two-loop massless ladder vertex, one off-shell leg, a single scale & $\geq 78$ \\
7 & central-mass double box & two loops, four massless legs, one massive central rung; three scales, seven letters & 109--112 \\
23 & massless-rung kite, four masses & two-loop kite with squared masses $(1,4,9,2,0)$; seven letters over four square roots & 81--110 \\
24 & non-planar hexa-box & two-loop five-point, off-shell legs $p_1^2=p_3^2=m^2$; 98 master integrals & 66--68 \\
\addlinespace[2pt]
\multicolumn{4}{@{}l}{\emph{Elliptic (maximal cut an elliptic curve)}}\\
8 & equal-mass sunrise & two loops, three equal masses, order $\eps^0$; the $\Gamma_1(6)$ curve & 140--159 \\
9 & sunrise $(1,1,2)$ & squared masses $(1,1,2)$, order $\eps^0$ & 159 \\
10 & sunrise $(1,2,3)$ & squared masses $(1,2,3)$, all three marked points distinct & 159 \\
11 & sunrise $(1,1,4)$ & squared masses $(1,1,4)$ & 159--289 \\
12 & sunrise at $\eps^1$ & the order-$\eps^1$ term for the three unequal-mass assignments & 288--308 \\
13 & equal-mass kite & the equal-mass sunrise with a massless rung across it; the $\Gamma_1(6)$ curve & 130--160 \\
14 & unequal-mass kite & non-modular sunrise sub-curve; solved by transport from $s=0$ & $\geq 125$ \\
15 & equal-mass ice-cream cone & two-loop three-point graph with four massive lines, order $\eps^0$ & 134 \\
16 & generic-mass ice-cream cone & squared masses $(1,2,3,5)$; non-modular family of curves & 118 \\
17 & ice-cream cone at $\eps^1$ & the weight-four order, on both mass slices & 134 \\
18 & $gg\to H$ elliptic masters & two-loop mixed QCD--electroweak; the two-mass sunrise curve $(m_t,m_t,M)$ & 48 \\
20 & $gg\to Z\gamma$ at exact $m_t$ & the pair of elliptic masters; boundary an Eichler integral on $\Gamma_0(4)$ & 109--112 \\
27 & light-by-light (sunrise) & three loops: the one-loop massive box with the equal-mass sunrise inserted on one rung & $\geq 45$ \\
28 & light-by-light (kite) & the massive box with the equal-mass kite inserted on one rung; dispersive & 59 \\
29 & light-by-light (QED parent) & the ten-propagator parent of the kite insertion, with two electron stubs & 50, 68 \\
33 & light-by-light ($N_f^2$) & the box with a vacuum-polarization insertion; dispersive, exact in $d$ & $\geq 37$ \\
35 & $q\bar q\to W^+W^-$ non-planar & two loops with exact top-quark mass, on a kinematic line; 76 masters, 23 new & 37--57 \\
\addlinespace[2pt]
\multicolumn{4}{@{}l}{\emph{K3 surface}}\\
19 & three-loop banana & four equal masses, order $\eps^0$; boundary constant $7\zeta_3$ & $\geq 110$ \\
31 & threshold banana (K3) & three loops, squared masses $(1,1,1,9)$; connection coefficient $c_{3/2}=-\sqrt3/(36\pi)$ & 195$^{b}$ \\
\addlinespace[2pt]
\multicolumn{4}{@{}l}{\emph{Calabi--Yau threefold}}\\
34 & four-loop banana & five equal masses, order $\eps^0$; operator AESZ~34 & 116 \\
32 & threshold banana (CY) & four loops, squared masses $(1,1,1,1,16)$; coefficients in the $\Gamma(\tfrac14)$ ring & 197$^{b}$ \\
\addlinespace[2pt]
\multicolumn{4}{@{}l}{\emph{Polylogarithmic top sector over a K3 sub-sector}}\\
21 & crossed light-by-light box & three loops: massive fermion loop, two crossed massless photon diagonals, four on-shell photons & 33--72 \\
\bottomrule
\end{tabular}
\end{table}

%% file: gallery15.tex
\begin{figure}[p]
\centering
\newsavebox{\gtilebox}%
\newlength{\gtileW}%
\newcommand{\gtile}[2]{%
  \begin{minipage}[b]{0.31\textwidth}\centering
    \begin{lrbox}{\gtilebox}#1\end{lrbox}%
    \setlength{\gtileW}{2.1cm*\ratio{\wd\gtilebox}{\ht\gtilebox+\dp\gtilebox}}%
    \ifdim\gtileW<0.93\linewidth
      \resizebox{!}{2.1cm}{\usebox{\gtilebox}}%
    \else
      \resizebox{0.93\linewidth}{!}{\usebox{\gtilebox}}%
    \fi\\[3pt]
    {\small #2}
  \end{minipage}}%
\gtile{\input{fig-kite.tex}}{two-mass kite; new}\hfill
\gtile{\input{fig-ggH.tex}}{$gg\to H$; known}\hfill
\gtile{\input{fig-sunrise-equalmass.tex}}{sunrise (elliptic); known}\\[14pt]
\gtile{\input{fig-c3-dbox.tex}}{central-mass double box; known}\hfill
\gtile{\input{fig-icc-generic.tex}}{generic-mass ice-cream cone (elliptic); new}\hfill
\gtile{\input{fig-lbl3e.tex}}{light-by-light (sunrise); new}\\[14pt]
\gtile{\input{fig-lbl3se.tex}}{light-by-light (kite); new}\hfill
\gtile{\input{fig-lbl3x.tex}}{light-by-light (crossed box, K3 sub-sector); new}\hfill
\gtile{\input{fig-lbl3kp.tex}}{light-by-light (QED parent); new}\\[14pt]
\gtile{\input{fig-lbl3vp.tex}}{light-by-light (vacuum polarization); new}\hfill
\gtile{\input{fig-zgamma.tex}}{$gg\to Z\gamma$, exact $m_t$; new}\hfill
\gtile{\input{fig-wpair-fc.tex}}{$q\bar q\to WW$ (non-planar); new}\\[14pt]
\gtile{\input{fig-banana-1119.tex}}{threshold banana (K3); new}\hfill
\gtile{\input{fig-banana-4l-11116.tex}}{threshold banana (CY$_3$); new}\hfill
\gtile{\input{fig-hexabox-eqmass.tex}}{non-planar hexa-box; new}
\caption{Twelve of the fifteen integrals drawn here, marked ``new'', have no
prior computation we could find; the three marked ``known'' ($gg\to H$, the equal-mass sunrise and the central-mass double box)
are literature integrals, with the published results used only at the
final comparison. Tile labels abbreviate the names in
Table~\ref{tab:portfolio} (the two-mass kite is its unequal-mass kite, the vacuum-polarization insertion
its $N_f^2$ row); elliptic, K3 and CY$_3$ (Calabi--Yau threefold) name the geometry of the maximal cut
where shown. Doubled or colored lines are massive, one color per distinct mass; thin black
lines are massless, and a colored wavy external leg is off shell; in the $gg\to H$ tile the dashed leg is the Higgs boson and the wavy internal line the electroweak boson of mass $M$.}
\label{fig:gallery15}
\end{figure}
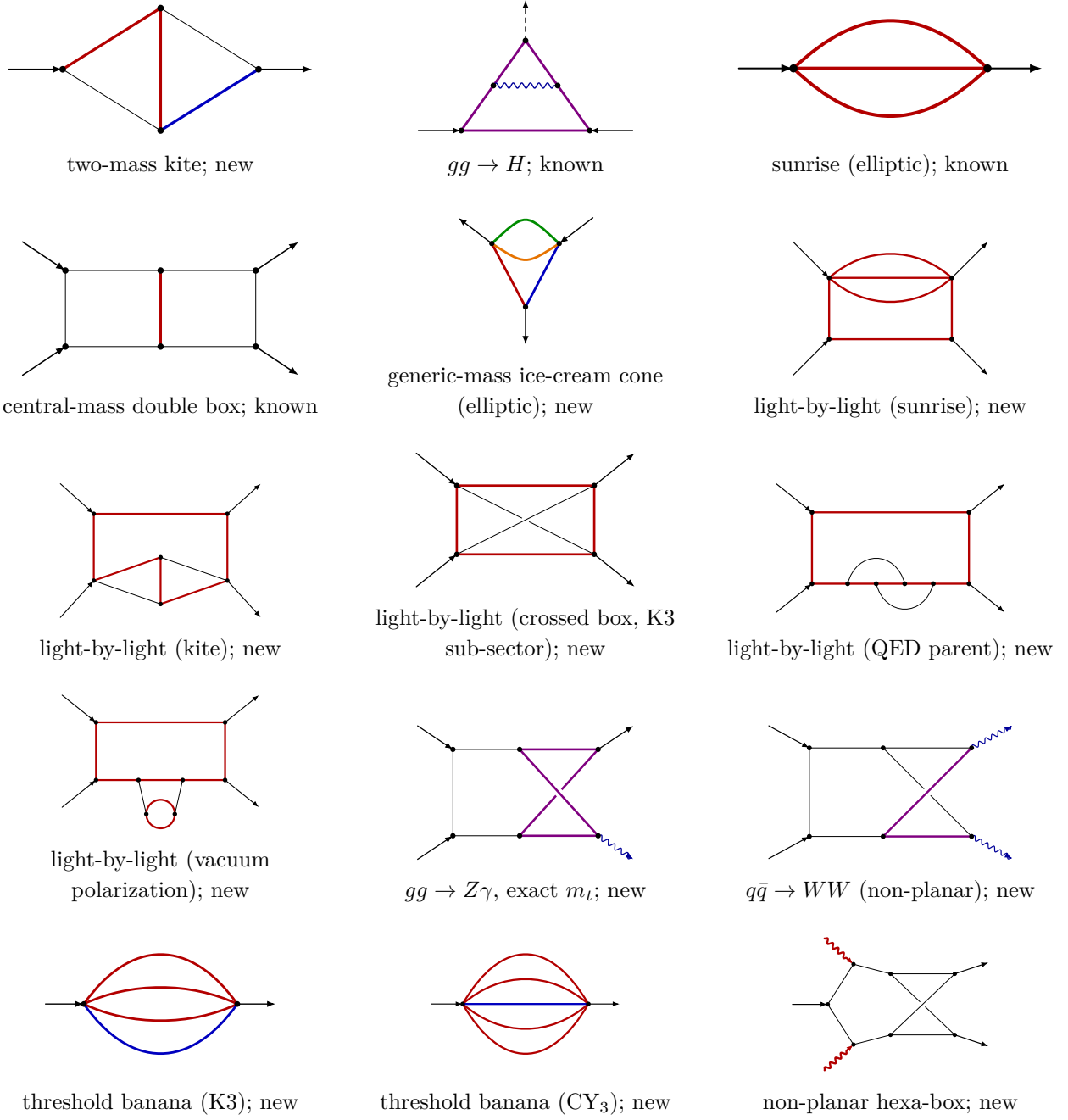

%% file: fig-kite.tex
\begin{tikzpicture}[fdiag]
  \coordinate (L) at (0,0);
  \coordinate (R) at (4.0,0);
  \coordinate (T) at (2.0,1.25);
  \coordinate (B) at (2.0,-1.25);
  \draw[ext] (-1.1,0) -- (L);
  \draw[ext] (R) -- (5.1,0);
  \draw[massM1]   (L) -- (T);   %
  \draw[massless] (T) -- (R);   %
  \draw[massless] (L) -- (B);   %
  \draw[massM2]   (B) -- (R);   %
  \draw[massM1]   (T) -- (B);   %
  \vertexdot{(L)} \vertexdot{(R)} \vertexdot{(T)} \vertexdot{(B)}
\end{tikzpicture}

%% file: fig-ggH.tex
\begin{tikzpicture}[fdiag]
  \coordinate (TL) at (0,0);
  \coordinate (TR) at (3.0,0);
  \coordinate (TT) at (1.5,2.1);
  \coordinate (GL) at (0.75,1.05);  %
  \coordinate (GR) at (2.25,1.05);  %
  \draw[massHeavy] (TL) -- (TR);
  \draw[massHeavy] (TL) -- (TT);
  \draw[massHeavy] (TR) -- (TT);
  \draw[thick, blue!60!black, decorate,
    decoration={snake, amplitude=0.6mm, segment length=2mm}] (GL) -- (GR);
  \draw[ext] (-1.0,0) -- (TL);
  \draw[ext] (4.0,0) -- (TR);
  \draw[thick,black,dashed,-{Latex[length=2mm]}] (TT) -- (1.5,3.05);
  \foreach \v in {(TL),(TR),(TT),(GL),(GR)} { \vertexdot{\v} }
\end{tikzpicture}

%% file: fig-sunrise-equalmass.tex
\begin{tikzpicture}[fdiag]
  \coordinate (L) at (0,0);
  \coordinate (R) at (3.2,0);
  \draw[ext] (-0.9,0) -- (L);
  \draw[ext] (R) -- (4.1,0);
  \draw[massM1] (L) .. controls (1.1,1.05) and (2.1,1.05) .. (R);
  \draw[massM1] (L) -- (R);
  \draw[massM1] (L) .. controls (1.1,-1.05) and (2.1,-1.05) .. (R);
  \vertexdot{(L)} \vertexdot{(R)}
\end{tikzpicture}

%% file: fig-c3-dbox.tex
\begin{tikzpicture}[fdiag]
  \coordinate (a1) at (0,0.8);
  \coordinate (a2) at (2.0,0.8);
  \coordinate (a3) at (4.0,0.8);
  \coordinate (b1) at (0,-0.8);
  \coordinate (b2) at (2.0,-0.8);
  \coordinate (b3) at (4.0,-0.8);
  \draw[massless] (a1) -- (a2) -- (a3);
  \draw[massless] (b1) -- (b2) -- (b3);
  \draw[massless] (a1) -- (b1);
  \draw[massless] (a3) -- (b3);
  \draw[massM1] (a2) -- (b2);
  \draw[ext] (-0.9,1.4) -- (a1);
  \draw[ext] (-0.9,-1.4) -- (b1);
  \draw[ext] (a3) -- (4.9,1.4);
  \draw[ext] (b3) -- (4.9,-1.4);
  \foreach \v in {(a1),(a2),(a3),(b1),(b2),(b3)} { \vertexdot{\v} }
\end{tikzpicture}

%% file: fig-icc-generic.tex
\begin{tikzpicture}[fdiag, scale=0.92]
  \coordinate (Vb) at (1.5,0.0);   %
  \coordinate (V1) at (0.65,1.6);  %
  \coordinate (V2) at (2.35,1.6);  %
  \draw[massM1] (Vb) -- (V1);                          %
  \draw[massM2] (Vb) -- (V2);                          %
  \draw[massM3] (V1) .. controls (1.5,2.40) .. (V2);   %
  \draw[massM4] (V1) .. controls (1.5,1.05) .. (V2);   %
  \draw[ext] (3.20,2.25) -- (V2);
  \draw[ext] (V1) -- (-0.20,2.25);
  \draw[ext] (Vb) -- (1.5,-0.95);
  \vertexdot{(Vb)} \vertexdot{(V1)} \vertexdot{(V2)}
\end{tikzpicture}

%% file: fig-lbl3e.tex
\begin{tikzpicture}[fdiag]
  \coordinate (TL) at (0,1.7);
  \coordinate (TR) at (3.4,1.7);
  \coordinate (BL) at (0,0);
  \coordinate (BR) at (3.4,0);
  \draw[massM1] (TR) -- (BR);
  \draw[massM1] (BR) -- (BL);
  \draw[massM1] (BL) -- (TL);
  \draw[massM1] (TL) -- (TR);                       %
  \draw[massM1] (TL) to[bend left=42]  (TR);        %
  \draw[massM1] (TL) to[bend right=42] (TR);        %
  \draw[ext] (-1.0,2.7) -- (TL);
  \draw[ext] (-1.0,-1.0) -- (BL);
  \draw[ext] (TR) -- (4.4,2.7);
  \draw[ext] (BR) -- (4.4,-1.0);
  \foreach \v in {(TL),(TR),(BL),(BR)} { \vertexdot{\v} }
\end{tikzpicture}

%% file: fig-lbl3se.tex
\begin{tikzpicture}[fdiag]
  \coordinate (TL) at (0,2.0);
  \coordinate (TR) at (4.0,2.0);
  \coordinate (BL) at (0,0);
  \coordinate (BR) at (4.0,0);
  \draw[massM1] (TL) -- (TR);
  \draw[massM1] (TL) -- (BL);
  \draw[massM1] (TR) -- (BR);
  \coordinate (Tk) at (2.0,0.70);
  \coordinate (Bk) at (2.0,-0.70);
  \draw[massM1]   (BL) -- (Tk);     %
  \draw[massless] (BL) -- (Bk);     %
  \draw[massless] (Tk) -- (BR);     %
  \draw[massM1]   (Bk) -- (BR);     %
  \draw[massM1]   (Tk) -- (Bk);     %
  \draw[ext] (-1.0,2.9) -- (TL);
  \draw[ext] (-1.0,-1.1) -- (BL);
  \draw[ext] (TR) -- (5.0,2.9);
  \draw[ext] (BR) -- (5.0,-1.1);
  \foreach \v in {(TL),(TR),(BL),(BR),(Tk),(Bk)} { \vertexdot{\v} }
\end{tikzpicture}

%% file: fig-lbl3x.tex
\begin{tikzpicture}[fdiag]
  \coordinate (TL) at (0,1.7);
  \coordinate (TR) at (3.4,1.7);
  \coordinate (BL) at (0,0);
  \coordinate (BR) at (3.4,0);
  \draw[massM1] (TL) -- (TR);
  \draw[massM1] (TR) -- (BR);
  \draw[massM1] (BR) -- (BL);
  \draw[massM1] (BL) -- (TL);
  \draw[massless] (TL) -- (BR);                 %
  \fill[white] (1.7,0.85) circle (2.8pt);       %
  \draw[massless] (BL) -- (TR);                 %
  \draw[ext] (-1.0,2.5) -- (TL);
  \draw[ext] (-1.0,-0.8) -- (BL);
  \draw[ext] (TR) -- (4.4,2.5);
  \draw[ext] (BR) -- (4.4,-0.8);
  \foreach \v in {(TL),(TR),(BL),(BR)} { \vertexdot{\v} }
\end{tikzpicture}

%% file: fig-lbl3kp.tex
\begin{tikzpicture}[fdiag]
  \coordinate (TL) at (0,2.0);
  \coordinate (TR) at (4.4,2.0);
  \coordinate (BL) at (0,0);
  \coordinate (BR) at (4.4,0);
  \coordinate (v1) at (1.0,0);
  \coordinate (v2) at (1.8,0);
  \coordinate (v3) at (2.6,0);
  \coordinate (v4) at (3.4,0);
  \draw[massM1] (BL) -- (TL);        %
  \draw[massM1] (TL) -- (TR);        %
  \draw[massM1] (TR) -- (BR);        %
  \draw[massM1] (BL) -- (v1);        %
  \draw[massM1] (v1) -- (v2);        %
  \draw[massM1] (v2) -- (v3);        %
  \draw[massM1] (v3) -- (v4);        %
  \draw[massM1] (v4) -- (BR);        %
  \draw[massless] (v1) .. controls +(0.25,0.95)  and +(-0.25,0.95)  .. (v3);  %
  \draw[massless] (v2) .. controls +(0.25,-0.95) and +(-0.25,-0.95) .. (v4);  %
  \draw[ext] (-1.0,-0.8) -- (BL);
  \draw[ext] (-1.0,2.8)  -- (TL);
  \draw[ext] (TR) -- (5.4,2.8);
  \draw[ext] (BR) -- (5.4,-0.8);
  \foreach \v in {(TL),(TR),(BL),(BR),(v1),(v2),(v3),(v4)} { \vertexdot{\v} }
\end{tikzpicture}

%% file: fig-lbl3vp.tex
\begin{tikzpicture}[fdiag]
  \coordinate (TL) at (0,1.7);
  \coordinate (TR) at (3.8,1.7);
  \coordinate (BL) at (0,0);
  \coordinate (BR) at (3.8,0);
  \coordinate (A) at (1.25,0);
  \coordinate (B) at (2.55,0);
  \coordinate (C) at (1.48,-1.0);   %
  \coordinate (D) at (2.32,-1.0);   %
  \draw[massM1] (BL) -- (TL);       %
  \draw[massM1] (TL) -- (TR);       %
  \draw[massM1] (TR) -- (BR);       %
  \draw[massM1] (BL) -- (A);        %
  \draw[massM1] (A) -- (B);         %
  \draw[massM1] (B) -- (BR);        %
  \draw[massless] (A) -- (C);       %
  \draw[massless] (D) -- (B);       %
  \draw[massM1] (1.9,-1.0) circle (0.42);
  \draw[ext] (-1.0,-0.8) -- (BL);
  \draw[ext] (-1.0,2.5) -- (TL);
  \draw[ext] (TR) -- (4.8,2.5);
  \draw[ext] (BR) -- (4.8,-0.8);
  \foreach \v in {(TL),(TR),(BL),(BR),(A),(B),(C),(D)} { \vertexdot{\v} }
\end{tikzpicture}

%% file: fig-zgamma.tex
\begin{tikzpicture}[fdiag]
  \coordinate (TL) at (0,1.1);   \coordinate (TM) at (1.7,1.1);   \coordinate (TR) at (3.7,1.1);
  \coordinate (BL) at (0,-1.1);  \coordinate (BM) at (1.7,-1.1);  \coordinate (BR) at (3.7,-1.1);
  \draw[massless] (TL) -- (TM);            %
  \draw[massless] (TL) -- (BL);            %
  \draw[massless] (BL) -- (BM);            %
  \draw[massHeavy] (TM) -- (TR);           %
  \draw[massHeavy] (BM) -- (BR);           %
  \draw[massHeavy] (TR) -- (BM);           %
  \draw[white, line width=6pt] ($(TM)!0.40!(BR)$) -- ($(TM)!0.60!(BR)$);
  \draw[massHeavy] (TM) -- (BR);           %
  \draw[ext] (-0.9,1.7) -- (TL);
  \draw[ext] (-0.9,-1.7) -- (BL);
  \draw[ext] (TR) -- (4.6,1.7);
  \draw[offshell] (BR) -- (4.6,-1.7);
  \vertexdot{(TL)}\vertexdot{(TM)}\vertexdot{(TR)}
  \vertexdot{(BL)}\vertexdot{(BM)}\vertexdot{(BR)}
\end{tikzpicture}

%% file: fig-wpair-fc.tex
\begin{tikzpicture}[fdiag]
  \coordinate (TL) at (0,1.2);   \coordinate (TM) at (2.0,1.2);   \coordinate (TR) at (4.4,1.2);
  \coordinate (BL) at (0,-1.2);  \coordinate (BM) at (2.0,-1.2);  \coordinate (BR) at (4.4,-1.2);
  \draw[massless] (TL) -- (TM);            %
  \draw[massless] (TM) -- (TR);            %
  \draw[massless] (BL) -- (BM);            %
  \draw[massless] (TL) -- (BL);            %
  \draw[massless] (TM) -- (BR);            %
  \draw[white, line width=7pt] ($(BM)!0.35!(TR)$) -- ($(BM)!0.65!(TR)$);
  \draw[massHeavy] (BM) -- (TR);           %
  \draw[massHeavy] (BM) -- (BR);           %
  \draw[ext] (-1.1,1.8) -- (TL);
  \draw[ext] (-1.1,-1.8) -- (BL);
  \draw[offshell] (TR) -- (5.5,1.8);
  \draw[offshell] (BR) -- (5.5,-1.8);
  \vertexdot{(TL)}\vertexdot{(TM)}\vertexdot{(TR)}
  \vertexdot{(BL)}\vertexdot{(BM)}\vertexdot{(BR)}
\end{tikzpicture}

%% file: fig-banana-1119.tex
\begin{tikzpicture}[fdiag]
  \coordinate (L) at (0,0);
  \coordinate (R) at (3.6,0);
  \draw[ext] (-0.9,0) -- (L);
  \draw[ext] (R) -- (4.5,0);
  \draw[massM1] (L) .. controls (1.1, 1.55) and (2.5, 1.55) .. (R);
  \draw[massM1] (L) .. controls (1.1, 0.52) and (2.5, 0.52) .. (R);
  \draw[massM1] (L) .. controls (1.1,-0.52) and (2.5,-0.52) .. (R);
  \draw[massM2] (L) .. controls (1.1,-1.55) and (2.5,-1.55) .. (R);
  \vertexdot{(L)} \vertexdot{(R)}
\end{tikzpicture}

%% file: fig-banana-4l-11116.tex
\begin{tikzpicture}[fdiag]
  \coordinate (L) at (0,0);
  \coordinate (R) at (3.6,0);
  \draw[ext] (-0.9,0) -- (L);
  \draw[ext] (R) -- (4.5,0);
  \draw[massM1] (L) .. controls (1.1, 1.9) and (2.5, 1.9) .. (R);
  \draw[massM1] (L) .. controls (1.1, 0.95) and (2.5, 0.95) .. (R);
  \draw[massM1] (L) .. controls (1.1,-0.95) and (2.5,-0.95) .. (R);
  \draw[massM1] (L) .. controls (1.1,-1.9) and (2.5,-1.9) .. (R);
  \draw[massM2] (L) -- (R);
  \vertexdot{(L)} \vertexdot{(R)}
\end{tikzpicture}

%% file: fig-hexabox-eqmass.tex
\begin{tikzpicture}[fdiag, scale=0.9]
  \coordinate (T7) at (1.6,0.95);    %
  \coordinate (B6) at (1.6,-0.95);   %
  \coordinate (V1) at (0.45,1.25);   %
  \coordinate (V2) at (-0.35,0);     %
  \coordinate (V3) at (0.45,-1.25);  %
  \coordinate (V4) at (3.6,-0.95);   %
  \coordinate (V5) at (3.6,0.95);    %
  \draw[massless] (T7) -- (V1);      %
  \draw[massless] (V1) -- (V2);      %
  \draw[massless] (V2) -- (V3);      %
  \draw[massless] (V3) -- (B6);      %
  \draw[massless] (T7) -- (V5);      %
  \draw[massless] (B6) -- (V4);      %
  \draw[massless] (T7) -- (V4);      %
  \draw[white, line width=5.5pt] ($(B6)!0.38!(V5)$) -- ($(B6)!0.62!(V5)$);
  \draw[massless] (B6) -- (V5);      %
  \draw[massM1, decorate, decoration={snake, amplitude=0.6mm, segment length=2mm, post length=1.8mm},
        -{Latex[length=2mm]}] (-0.45,2.05) -- (V1);
  \draw[massM1, decorate, decoration={snake, amplitude=0.6mm, segment length=2mm, post length=1.8mm},
        -{Latex[length=2mm]}] (-0.45,-2.05) -- (V3);
  \draw[ext] (-1.45,0) -- (V2);
  \draw[ext] (V5) -- (4.65,1.30);
  \draw[ext] (V4) -- (4.65,-1.30);
  \foreach \v in {(T7),(B6),(V1),(V2),(V3),(V4),(V5)} { \vertexdot{\v} }
\end{tikzpicture}

%% file: s4_integrals.tex
\newpage

\section{Evidence integrals\label{sec:evidence}}
In this section, we explain how the Bayesian evidence of a polynomial model for count data is an integral of the same kind as for a Feynman diagram, so that it can be evaluated with the same tools (Section~\ref{sec:amplitudes}). For the applications below (Table~\ref{tab:types}) the findings are in the papers cited.
\subsection{Bayesian evidence integrals\label{sec:quartet}}

Bayesian model selection ranks competing hypotheses by their evidence, or marginal likelihood: the probability of the data under a hypothesis, integrated over its parameters against their prior~\cite{Kass1995}. In general a model $M$ with parameters $\theta$ assigns the data $D$ a likelihood $P(D\,|\,\theta,M)$, and a prior density $\pi(\theta\,|\,M)$ weights the parameters. Bayes' theorem gives their posterior distribution,
\begin{equation}
P(\theta\,|\,D,M)=\frac{P(D\,|\,\theta,M)\,\pi(\theta\,|\,M)}{Z_M}
\label{eq:bayes}
\end{equation}
with normalization
\begin{equation}
  Z_M=\int P(D\,|\,\theta,M)\,\pi(\theta\,|\,M)\,d\theta\,,
\end{equation}
called the {\it evidence} or the {\it marginal likelihood}. The evidence allows us to compare models: the posterior odds of $M_1$ over $M_2$ are the prior odds times the Bayes factor $Z_{M_1}/Z_{M_2}$. As a toy example, toss a coin $n$ times and observe $k$ heads. Under the hypothesis of a fair coin there is no parameter and $Z_{\rm fair}=2^{-n}$. Under the hypothesis of an unknown bias $x$, the probability of heads, with a uniform prior on $[0,1]$,
\begin{equation}
Z_{\rm bias}=\int_0^1 x^{k}(1-x)^{\,n-k}\,dx=\frac{k!\,(n-k)!}{(n+1)!}\,.
\label{eq:coin}
\end{equation}
Note that the unknown bias is integrated over: any bias is allowed with a flat prior on $x$. If the data come out as $k=8$ heads after $n=10$ flips, this gives $Z_{\rm fair}=1/1024$ and $Z_{\rm bias}=1/495$, a Bayes factor $Z_{\rm bias}/Z_{\rm fair}\approx 2.1$ in favor of a biased coin. In words, the observed eight heads in ten flips is 2.1 times as probable under the biased-coin hypothesis as under the fair-coin hypothesis. More technically, even prior odds become posterior odds of around $2:1$, so that the probability is $2.1/(1+2.1)\approx 2/3$ that the coin is biased. A common convention~\cite{Kass1995} is that a Bayes factor below about three is weak evidence and ``not worth more than a bare mention''.

The integrand of Eq.~\eqref{eq:coin} is a product of the outcome probabilities, $x$ for heads and $1-x$ for tails, each raised to the number of times that outcome occurred. The same structure appears whenever the data are counts and the probability of each outcome is a polynomial in the parameters. With outcome $k$ observed $u_k$ times, outcome probabilities $P_k(x)$ in parameters $x=(x_1,\dots,x_d)$ scaled to the unit cube, and a prior proportional to $\prod_e x_e^{\alpha_e-1}$ (uniform for $\alpha_e=1$), the evidence is
\begin{equation}
Z \;=\; \int_0^{1}\!\cdots\!\int_0^{1} \; \prod_k P_k(x)^{u_k} \, \prod_e x_e^{\alpha_e-1} \, dx\,,
\label{eq:evidence}
\end{equation}
with the polynomials $P_k$ raised to the data counts $u_k$ and the exponents $\alpha_e$ set by the prior. In most models of interest it has no elementary closed form. Equation~\eqref{eq:evidence} is a generalized Euler integral of the class of Eq.~\eqref{eq:eulerclass}~\cite{GKZ1990}, and the counts $u_k$ occupy the place of the propagator powers in the Lee--Pomeransky representation of a Feynman integral~\cite{LeePomeransky}. Algebraic statisticians have written statistical models in this polynomial form~\cite{SturmfelsSullivant2005} and computed such integrals exactly for small models~\cite{LinSturmfelsXu2009}. We apply the reduction, recurrence and finite-field routines built for Feynman integrals to Eq.~\eqref{eq:evidence} unchanged.

\subsection{Phylogenetics\label{sec:phylo}}
\begin{figure}[t]
\centering
{\footnotesize (a)}\par\smallskip
\input{fig-phylo-quartet.tex}\par\bigskip
{\footnotesize (b)}\par\smallskip
\input{fig-phylo-alignment.tex}
\caption{The evidence of a four-taxon tree depends on the DNA alignment only through the counts $N_p$ of site-pattern classes, which occupy in Eq.~\eqref{eq:phyloevidence} the place of the propagator powers in a Feynman integral.
(a)~The rooted family tree of human, chimpanzee, gorilla and orangutan (internal nodes: the chimpanzee--human and human--gorilla last common ancestors, CHLCA and HGLCA) and the same tree unrooted, the quartet topology $12|34$ with pendant edge variables $x_1,\dots,x_4$ and internal $x_5$, where $x_e=e^{-4t_e/3}$; unrooting leaves the root as a tick mark on the orangutan branch.
(b)~A short alignment, one column per site; the shaded columns show the pattern class supporting $12|34$ against the rivals $13|24$ and $14|23$.}
\label{fig:phylo}
\end{figure}
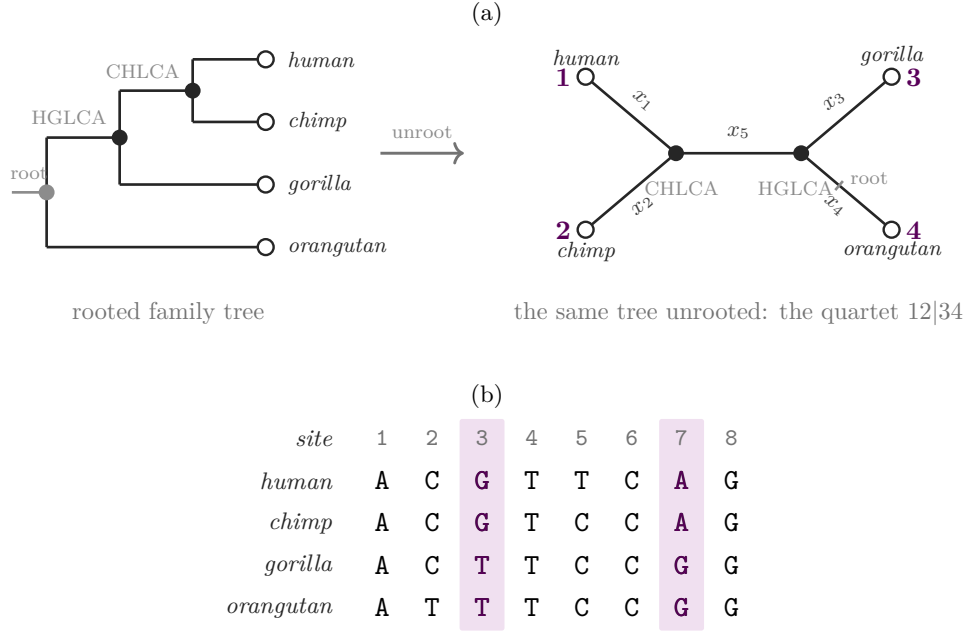

Phylogenetics offers a concrete example. The problem is to infer, from the DNA of several living species, the tree by which they descend from a common ancestor: its branching order (the topology) and the lengths of its branches. The data are aligned sequences, one row per species and one column per site, so that each column records which of the four bases every species carries at one position (Fig.~\ref{fig:phylo}b). A substitution model, a Markov chain on the four bases running along the branches, gives the probability of each column as a function of the branch lengths, and the likelihood of the alignment is the product of these probabilities over the columns. Each topology is then a hypothesis $M$ in the sense of Eq.~\eqref{eq:bayes} with the branch lengths as its parameters $\theta$, and rival topologies are compared by their evidences. For thirty years phylogeneticists have estimated these evidences by Monte Carlo (thermodynamic integration, stepping-stone sampling, nested sampling), with error bars that are themselves estimates~\cite{LartillotPhilippe2006,Xie2011,Fourment2020}; packages such as MrBayes~\cite{MrBayes2012} work this way.

We make the correspondence between evidence integrals and Feynman integrals explicit for the smallest phylogenetic problem. Its hypotheses are the three tree topologies (branching orders) that could relate four species (a quartet), and its parameters are the five branch lengths of the unrooted tree $\tau$ (Fig.~\ref{fig:phylo}a). The alignment has $N$ sites. We take the Jukes--Cantor model, the simplest base-symmetric substitution model: each site evolves independently along $\tau$ by a continuous-time Markov chain on the four bases with all substitutions equally likely. Along an edge of length $t$ (expected substitutions per site) a base survives or changes with probabilities
\begin{equation}
P_{\rm same}=\frac{1+3x}{4}\,,\qquad
P_{\rm diff}=\frac{1-x}{4}\,,\qquad
x=e^{-4t/3}\,.
\label{eq:jc}
\end{equation}

Because the Jukes--Cantor model treats the bases alike, the patterns of bases that a site can show across the four species fall into classes $p$, the patterns within a class having equal probability, with $N_p$ sites in class $p$ (Fig.~\ref{fig:phylo}b). The probability $f_p$ of a pattern in class $p$ is a sum, over the states of the two internal nodes, of products of five edge factors, each linear in one $x_e$ by Eq.~\eqref{eq:jc}. Thus $f_p$ is a polynomial with rational coefficients, multilinear in $x_1,\dots,x_5$, and the likelihood of the alignment is $\prod_p f_p^{N_p}$. We put an independent exponential prior of rate $4/3$ on each branch length. Its density cancels the Jacobian of $t\mapsto x$ exactly, so that each $x_e$ is uniformly distributed on $[0,1]$ and the evidence for the topology is
\begin{equation}
Z_\tau \;=\; \int_{[0,1]^d} \prod_p f_p(x)^{N_p}\,d^dx\,,
\label{eq:phyloevidence}
\end{equation}
with $d=5$ for a quartet and $d=7$ for a quintet. Hence Eq.~\eqref{eq:phyloevidence} is Eq.~\eqref{eq:evidence} with $P_k=f_p$, $u_k=N_p$ and every $\alpha_e=1$. Each of the three topologies has its own pattern polynomials, and the fifteen topologies of a quintet are scored likewise.

For four- and five-taxon trees under the standard base-symmetric substitution models with exponential branch-length priors, the evidence is an exact rational number, a ratio of explicit integers~\cite{PhyloNote,JackandjillNote}. It is rational because each monomial $c_m x^m$ of $\prod_p f_p^{N_p}$ integrates over the unit cube to $c_m\prod_e 1/(m_e+1)$; by multilinearity every exponent $m_e$ is at most $N=\sum_p N_p$, which bounds the denominator of $Z_\tau$ in terms of the counts alone. We evaluate the monomial sum modulo machine-word primes and reconstruct the rational by Chinese remaindering~\cite{PhyloNote}; two reconstructions from disjoint sets of primes must then agree. It is also compared with an independent arbitrary-precision quadrature of Eq.~\eqref{eq:phyloevidence}.

As with a family of Feynman integrals, finitely many master integrals span the family of integrals $Z_\tau$, and members whose counts differ by integers are connected by linear contiguity recurrences with rational coefficients, the counterpart of the integration-by-parts identities. The one-parameter case is classical, and the coin of Eq.~\eqref{eq:coin} is its simplest instance. For $k$ successes in $n$ Bernoulli trials under a $\mathrm{Beta}(a,b)$ prior on the success rate, the evidence $Z(a,b;k,n)$ obeys
\begin{equation}
\frac{Z(a,b;\,k+1,\,n+1)}{Z(a,b;\,k,\,n)}\;=\;\frac{a+k}{a+b+n}\,,
\label{eq:succession}
\end{equation}
which is the posterior predictive probability of the next success (Laplace's rule of succession). Read as a computation, Eq.~\eqref{eq:succession} and its counterpart for a failure, with $b+n-k$ in the numerator, give the evidence of any dataset from $Z(\text{no data})=1$ in $n$ exact rational multiplications. For the quartet the recurrence has higher order and matrix coefficients. We obtain it with Annihilator (Table~\ref{tab:toolkit}), the routine written to find the Picard--Fuchs operators of banana-graph periods, used unchanged with shifts of pattern counts replacing shifts of propagator powers~\cite{PhyloNote}.

Two topologies can have exactly equal evidence. If a relabeling of the taxa exchanges two rival topologies while preserving the pattern counts, their evidences are equal by symmetry. The exact computation then returns identical rational numbers, whereas a floating-point evaluator gives two nearby numbers and cannot report the tie, which no amount of sampling could establish.

Outside the base-symmetric models with exponential priors (for richer substitution models, or with rate variation across sites) we compute the evidence integral as an enclosure proven to contain it, in the ball arithmetic of Section~\ref{sec:ball}. A comparison of two hypotheses then becomes an inequality between intervals, which is conclusive only when the lower bound on one evidence exceeds the upper bound on the other and is otherwise reported as inconclusive. For the exact computation and the enclosure alike, the practical limit is cost, which rises steeply with the number of species and of rate classes. The package JaCK \& Jill~\cite{JackandjillNote,PhyloNote} returns these evidences for four species, exactly under the Jukes--Cantor and Kimura models and as proven enclosures otherwise.

\subsection{Evidence integrals in other fields\label{sec:areas}}

Integrals of the form of Eq.~\eqref{eq:evidence} and the recurrences that connect them arise in other fields where a probabilistic model of counts is compared with data. In the statistics of mixtures one asks whether data that look mixed come from one population or several. For a discrete variable the marginal likelihoods of models with different numbers of components, treated as intractable for decades, are again integrals of polynomials raised to the data counts. The BootLoops tool Mixalot~\cite{MixturesNote} evaluates them as finite sums by the method of Ref.~\cite{LinSturmfelsXu2009}. Shared authorship is also a mixture problem. Counts of function words (articles, prepositions and the like) have mostly been used to assign a disputed text to one author. Treating a text as a mixture makes each proposed division among authors a hypothesis with its own evidence in closed form~\cite{AuthorsNote}.

Evidence integrals arise in several other fields as well. In neutral ecology the likelihood is the probability of the observed species abundances on a forest plot under Hubbell's model~\cite{Hubbell2001,Etienne2005}. It obeys an exact recurrence in the abundance counts, although it is usually approximated by Monte Carlo simulation. In single-cell biology the rates and sizes of transcriptional bursts (brief episodes of RNA synthesis separated by long silent intervals) are fitted to RNA counts per cell~\cite{Larsson2019}. The likelihood is a product of hundreds of factors, and standard floating-point tools give no bound on the accumulated rounding error of the product~\cite{GeneburstingNote}. In population genetics the site-frequency spectrum of a selected allele (how many variants appear on one, two or more sampled chromosomes) underlies the inference of fitness effects. The spectrum predicted by the standard theory (the Sawyer--Hartl Poisson random field) could not be evaluated accurately at strong selection and large sample sizes, where floating-point arithmetic fails~\cite{MutationflowNote}. We compute the bursting likelihood in ball arithmetic and the site-frequency spectrum in exact rational arithmetic.

%% file: fig-phylo-quartet.tex
\begin{tikzpicture}[
  scale=0.92, transform shape,
  edge/.style  = {line width=1pt, draw=black!85},
  leaf/.style  = {circle, draw=black!85, fill=white, line width=0.9pt, inner sep=0pt, minimum size=6.5pt},
  inode/.style = {circle, fill=black!85, inner sep=0pt, minimum size=6.5pt},
  taxon/.style = {font=\bfseries, text=violet!70!black},
  sp/.style    = {font=\itshape\footnotesize, text=black!85},
  anc/.style   = {font=\scriptsize, text=black!45},
  elab/.style  = {font=\footnotesize, text=black!75},
  lab/.style   = {font=\small, text=black!55},
]
  \begin{scope}[shift={(-8.4,0)}]
    \coordinate (rH) at (1.6, 1.35); \coordinate (rC) at (1.6, 0.45);
    \coordinate (rG) at (1.6,-0.45); \coordinate (rO) at (1.6,-1.35);
    \coordinate (rA) at ( 0.55, 0.90);   %
    \coordinate (rB) at (-0.50, 0.225);  %
    \coordinate (rR) at (-1.55,-0.5625); %
    \draw[edge] (0.55,1.35)--(0.55,0.45);   \draw[edge] (0.55,1.35)--(rH);  \draw[edge] (0.55,0.45)--(rC);
    \draw[edge] (-0.50,0.90)--(-0.50,-0.45); \draw[edge] (-0.50,0.90)--(rA); \draw[edge] (-0.50,-0.45)--(rG);
    \draw[edge] (-1.55,0.225)--(-1.55,-1.35); \draw[edge] (-1.55,0.225)--(rB); \draw[edge] (-1.55,-1.35)--(rO);
    \draw[edge, draw=black!45] (-2.05,-0.5625)--(rR);
    \node[inode] at (rA) {}; \node[inode] at (rB) {};
    \node[inode, fill=black!45] at (rR) {};
    \node[anc, anchor=south] at (-1.80,-0.50) {root};
    \node[anc, anchor=south east] at (0.50,0.95) {CHLCA};
    \node[anc, anchor=south east] at (-0.55,0.28) {HGLCA};
    \node[leaf] at (rH) {}; \node[leaf] at (rC) {};
    \node[leaf] at (rG) {}; \node[leaf] at (rO) {};
    \node[sp, anchor=west] at (1.78, 1.35) {human};
    \node[sp, anchor=west] at (1.78, 0.45) {chimp};
    \node[sp, anchor=west] at (1.78,-0.45) {gorilla};
    \node[sp, anchor=west] at (1.78,-1.35) {orangutan};
    \node[lab] at (0.2,-2.30) {rooted family tree};
  \end{scope}
  \draw[->, line width=1.1pt, black!55] (-5.15,0)--(-3.95,0);
  \node[anc] at (-4.55,0.30) {unroot};
  \coordinate (L) at (-0.9,0); \coordinate (R) at (0.9,0);
  \coordinate (t1) at (-2.2, 1.1); \coordinate (t2) at (-2.2,-1.1);
  \coordinate (t3) at ( 2.2, 1.1); \coordinate (t4) at ( 2.2,-1.1);
  \draw[edge] (t1)--(L) node[elab,midway,above=1pt,sloped]{$x_1$};
  \draw[edge] (t2)--(L) node[elab,midway,below=1pt,sloped]{$x_2$};
  \draw[edge] (R)--(t3) node[elab,midway,above=1pt,sloped]{$x_3$};
  \draw[edge] (R)--(t4) node[elab,midway,below=1pt,sloped]{$x_4$};
  \draw[edge] (L)--(R)  node[elab,midway,above=2pt]{$x_5$};
  \draw[black!45, line width=1pt] (1.39,-0.53)--(1.50,-0.40);
  \node[anc] at (1.88,-0.36) {root};
  \node[inode] at (L) {}; \node[inode] at (R) {};
  \node[anc] at (-0.82,-0.50) {CHLCA};
  \node[anc] at ( 0.82,-0.50) {HGLCA};
  \node[leaf] at (t1) {}; \node[leaf] at (t2) {};
  \node[leaf] at (t3) {}; \node[leaf] at (t4) {};
  \node[taxon] at (-2.52, 1.1){1}; \node[taxon] at (-2.52,-1.1){2};
  \node[taxon] at ( 2.52, 1.1){3}; \node[taxon] at ( 2.52,-1.1){4};
  \node[sp,above=1pt] at (t1){human};   \node[sp,below=1pt] at (t2){chimp};
  \node[sp,above=1pt] at (t3){gorilla}; \node[sp,below=1pt] at (t4){orangutan};
  \node[lab] at (0,-2.30) {the same tree unrooted: the quartet $12|34$};
\end{tikzpicture}

%% file: fig-phylo-alignment.tex
\begin{tikzpicture}[x=0.66cm, y=0.56cm,
  lt/.style = {font=\ttfamily},
  hd/.style = {font=\ttfamily\footnotesize, text=black!55},
  rw/.style = {font=\itshape\footnotesize, text=black!85, anchor=east},
  sig/.style= {font=\ttfamily, text=violet!70!black},
]
  \fill[violet!12, rounded corners=1pt] (2.55,0.45) rectangle (3.45,-4.55);
  \fill[violet!12, rounded corners=1pt] (6.55,0.45) rectangle (7.45,-4.55);
  \node[rw] at (0.2,0) {site};
  \foreach \x in {1,...,8} {\node[hd] at (\x,0) {\x};}
  \node[rw] at (0.2,-1) {human};     \foreach \x/\c in {1/A,2/C,3/G,4/T,5/T,6/C,7/A,8/G}{\node[lt] at (\x,-1){\c};}
  \node[rw] at (0.2,-2) {chimp};     \foreach \x/\c in {1/A,2/C,3/G,4/T,5/C,6/C,7/A,8/G}{\node[lt] at (\x,-2){\c};}
  \node[rw] at (0.2,-3) {gorilla};   \foreach \x/\c in {1/A,2/C,3/T,4/T,5/C,6/C,7/G,8/G}{\node[lt] at (\x,-3){\c};}
  \node[rw] at (0.2,-4) {orangutan}; \foreach \x/\c in {1/A,2/T,3/T,4/T,5/C,6/C,7/G,8/G}{\node[lt] at (\x,-4){\c};}
  \foreach \y/\c in {-1/G,-2/G,-3/T,-4/T}{\node[sig] at (3,\y){\c};}
  \foreach \y/\c in {-1/A,-2/A,-3/G,-4/G}{\node[sig] at (7,\y){\c};}
\end{tikzpicture}

%% file: s5_enumeration.tex
\section{Exhaustive enumeration and proof}\label{sec:enumeration}

Some quantitative questions are settled only by examining every case, and others only by a proof that no case of the sought kind exists. Exhaustive search with a proof of completeness has been part of pure mathematics since the four-color theorem of 1976~\cite{AppelHaken1976,AppelHaken1977}. Besides exhaustive search we use creative telescoping~\cite{Petkovsek1996,Koutschan2013}, a computer-algebra method for proving that an integral or sum satisfies a given differential equation or recurrence, and dynamic programming~\cite{Bellman1957}, which finds exact optima over ordered partitions.

\subsection{Complete censuses and planted instances}\label{sec:complete}
A finite enumeration (a census) can only be considered rigorously complete if it comes with a proof that it missed nothing. Such a proof is an identity that the full list satisfies and a partial list cannot. For integral lattices the identity is Siegel's mass formula~\cite{Siegel1935}, which fixes in advance a weighted count of all lattice classes of a given local type (a genus). An enumeration is complete exactly when the weights of the classes found add up to the mass. As an additional check, if a second, independent enumeration algorithm exists the two lists must coincide.

A claim that no solution exists needs a proof of a different kind, a theorem rather than a checked list (Section~\ref{sec:impossible}). Enumerations and nonexistence results are rigorous as mathematics: completeness is proved, not sampled, and nothing can be missed if the program does what the proof says. However, when AI is doing the proving, there is another failure mode: bugs in the code. A computer-assisted proof is only as reliable as the code that executes it. A program that silently skips part of a search space, not because it didn't know that space was there, but because of a bug in the code, may mistakenly claim it has been exhaustive. Thus it is important to have controls so that this doesn't happen. One powerful way to do this is to plant synthetic targets of the same kind and size and make sure the solver finds them. The planted control tests the implementation, not the argument; this is a standard precaution of computer-assisted proofs. Note that a heuristic search gives no such guarantee even with correct code: a Markov chain Monte Carlo sampler can pass its convergence diagnostics while missing a well-separated mode, which is one reason we enumerate rather than sample where enumeration is possible.

For a concrete example, consider string compactifications with flux. There, the allowed fluxes form a lattice (each point a candidate vacuum), and such vacua are usually explored one at a time~\cite{StringvacNote}. To answer a question about all of them at once (whether any allowed flux stabilizes every modulus, for example) we need the complete list. Here the completeness identity is again Siegel's mass formula, and a BootLoops routine Terrier, named after the string landscape, can carry out the enumeration in exact lattice arithmetic. %
Another example concerns the mysterious Voynich manuscript. Readings of this manuscript can be written as probability models, fitted on part of the pages and judged on the rest~\cite{VoynichNote}. A failure on the withheld pages excludes a reading only if the test has also been shown to detect disguised real text of the same kind (the planted-instance rule).

An important application is to the operating characteristics of adaptive and Bayesian Phase~II trial designs. The type~I error rate and power are usually estimated from a few thousand simulated trials, each with Monte Carlo noise~\cite{HealthtrialsNote}. The actual decision rule can often be stated as requiring either finite sums over outcome sequences or low-dimensional integrals. These can then be computed exactly or as proven enclosures.

A final application is to Medicare's star-rating systems, which grade one-dimensional quality scores by cut points, defined as the boundaries of the bins that minimize the within-bin sum of squared deviations. In practice the bins are computed with clustering heuristics that may stop short of that minimum~\cite{HealthstarsNote}. In one dimension, however, the optimal bins are intervals, and the exact minimum has been computable since 1958 by a dynamic program over the cut points~\cite{Fisher1958}. That dynamic program is an exhaustive search over every placement of the cut points, organized so that it runs in polynomial time, and it returns the global minimum together with the proof that no other placement does better; a clustering heuristic returns a local minimum and no such statement.

\subsection{Impossibility as a theorem}\label{sec:impossible}
Exhaustive enumeration can also help BootLoops compute integrals, particularly by the constraint leg of the bootstrap (Section~\ref{sec:method-pipeline}). It may happen that in exhausting a class of candidate formulas for a given integral one finds, instead of a list, a proof that no member of the class equals the integral. This happened in the computation of the anisotropic Watson integral,
\begin{equation}
W(a,b,c;w) \;=\; \frac{1}{\pi^3}\int_0^\pi\!\!\int_0^\pi\!\!\int_0^\pi
\frac{d\theta_1\, d\theta_2\, d\theta_3}{w - a\cos\theta_1 - b\cos\theta_2 - c\cos\theta_3}\,,
\label{eq:watson}
\end{equation}
the generating function of the return probabilities of a random walk on the simple cubic lattice that hops along the three axes with rates $(a,b,c)$. Whether a closed form exists at generic rates has been an open question since Watson evaluated the isotropic case $a=b=c$ in 1939~\cite{Watson1939}. Every previously known exact result lies on a locus where two rates coincide, on which the integral factors into two complete elliptic integrals~\cite{DelvesJoyce2001,Zucker2011}.

The integral of Eq.~\eqref{eq:watson} is holonomic in $w$ (it satisfies a linear differential equation with polynomial coefficients). The kinds of closed form it can have depend on the order and the reducibility of the minimal such equation. From the series coefficients of $W$ we obtained, in exact arithmetic, a candidate annihilating operator of order~5. We then proved by creative telescoping that it annihilates $W$: applied to the integrand, the operator gives an exact total derivative in the angles~\cite{Koutschan2013,Kauers2011,WatsonNote}. For both steps, and for the proof that no operator of lower order exists, we again used Annihilator (Table~\ref{tab:toolkit}), the routine written for Feynman integrals and reused above for phylogenetic evidence.

The impossibility argument then runs as follows. A complete elliptic integral satisfies a differential equation of order~2, and any candidate of the sought form (a product of two complete elliptic integrals with an algebraic prefactor) therefore satisfies an equation of order~4. The minimal operator of $W$ has order~5 and is irreducible for general rates: its differential Galois group is the full $SO(5)$, which excludes every function built from solutions of second-order equations. No product of complete elliptic integrals and no ${}_3F_2$ represents the generic Watson integral. At one specially constructed rate tuple with complex multiplication the integral has a closed form in $\Gamma(k/16)$ values, and for general rates the integral is a period of a K3 surface~\cite{WatsonNote}.

Thus the nonexistence of the sought closed form is a theorem about the orders of differential operators, and its proof is produced by computer algebra. The telescoping identity on which the proof rests can be confirmed by differentiation alone, without the language model. A class of closed forms can also be excluded by a linear-independence obstruction, when the differential equation requires a function that no combination of members of the class can equal, as for the crossed light-by-light box of Section~\ref{sec:smatrix}, whose top-sector cut is polylogarithmic but which has no polylogarithmic form over its own letters because of a K3 sub-sector (Table~\ref{tab:portfolio}).

%% file: s6_ball.tex
\section{Ball arithmetic and audits\label{sec:ball}}

Many calculations in quantitative science end in a number computed in floating-point arithmetic, with a rounding error that is neither estimated nor propagated. Rigorous alternatives have existed since the interval analysis of the 1960s~\cite{Moore2009}. In validated numerics, bounds on rounding error are carried through entire computations and into computer-assisted proofs~\cite{Tucker2011,Rump2010}. The Arb library~\cite{Johansson2017} made interval arithmetic in ball form (a midpoint with a radius, in place of two endpoints) efficient at arbitrary precision. We use ball arithmetic to compute numbers with proven error bounds and to examine numbers already in print.

\subsection{Balls and Muller's recurrence\label{sec:muller}}

A ball is a midpoint--radius pair $m\pm r$. It denotes the interval $[m-r,\,m+r]$, and an arithmetic operation $\circ$ on two balls is defined to return a ball that contains the exact result for every choice of one member from each ball,
\begin{equation}
x\in[m_1\pm r_1],\ \ y\in[m_2\pm r_2]\ \ \Longrightarrow\ \ x\circ y\,\in\,[m_1\pm r_1]\circ[m_2\pm r_2]\,.
\label{eq:ball}
\end{equation}
Elementary and special functions extend to balls in the same way~\cite{Moore2009,Johansson2017}. Every operation on balls rounds the radius outward, which keeps the true value within $r$ of $m$ through any sequence of operations; $\log_{10}(|m|/r)$ is then the number of guaranteed digits. Our implementation, Baller~\cite{BallerNote}, is built on Arb; given a requested number of digits, it raises the working precision until that many digits are guaranteed, and otherwise returns a refusal naming the test that failed.

\begin{figure}[t]
\centering
\includegraphics[width=\textwidth]
{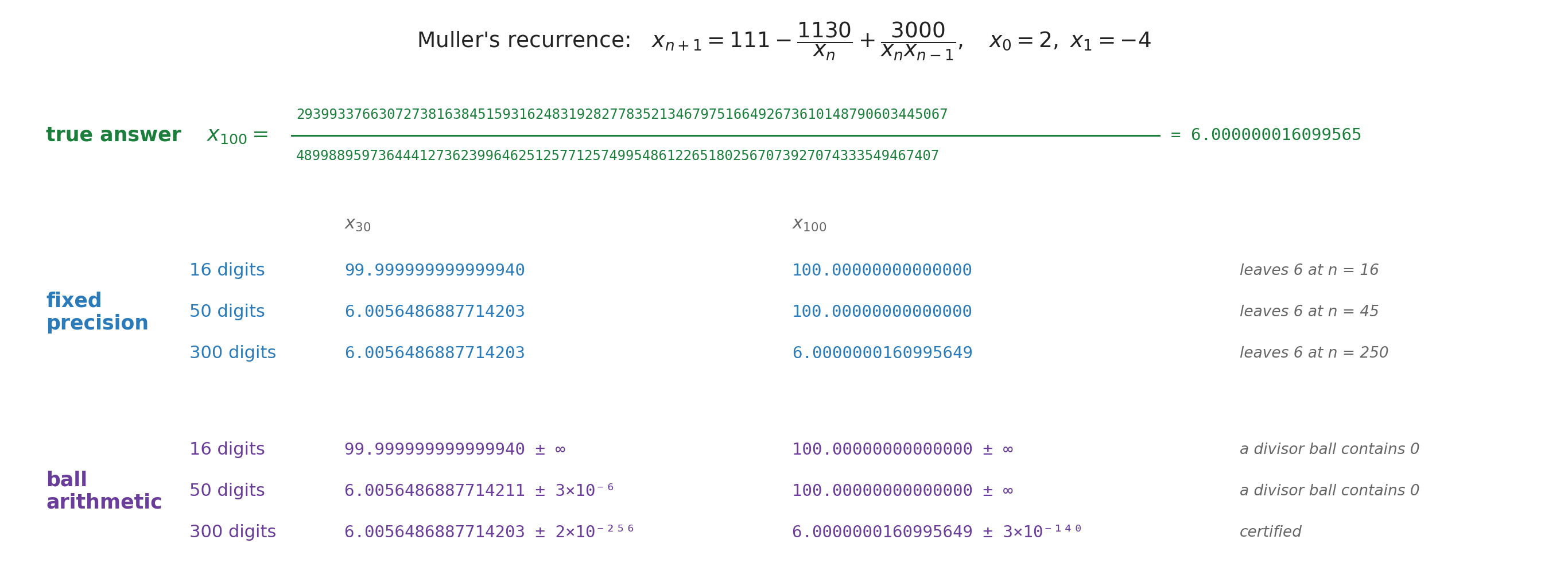}
\caption{At 16 and 50 digits of working precision, fixed-precision arithmetic returns a wrong value of $x_{100}$ in Muller's recurrence, Eq.~\eqref{eq:muller}, with no indication that it is wrong, whereas ball arithmetic returns either certified digits or an infinite radius.
Top: the recurrence with its seeds, and the exact $x_{100}$, a ratio of a 79-digit integer to a 78-digit integer, approximately $6.000000016099565$.
Middle: $x_{30}$ and $x_{100}$ computed in fixed precision at 16, 50 and 300 digits.
At 16 and 50 digits $x_{100}$ prints as exactly 100, because rounding error excites a second solution of the recurrence that grows about seventeen-fold per step; at 300 digits the value is correct.
The notes at right give the step at which each fixed-precision iterate leaves the neighborhood of 6.
Bottom: the same three computations in ball arithmetic.
Where a divisor ball contains zero the radius is infinite and no digit is asserted; at 300 digits every printed digit of $x_{100}$ is guaranteed.
The midpoints in the bottom rows track the fixed-precision values because a ball's midpoint is computed by the same floating-point operations; only the radius indicates which of its digits are established.}
\label{fig:muller}
\end{figure}

We illustrate the role of the radius with Muller's recurrence, a standard example in numerical analysis. The sequence
\begin{equation}
x_{n+1}=111-\frac{1130}{x_n}+\frac{3000}{x_n\,x_{n-1}}\,,\qquad x_0=2,\quad x_1=-4,
\label{eq:muller}
\end{equation}
converges to 6, and the true $x_{100}$ is an exact fraction of 79- and 78-digit integers, approximately $6.000000016099565$ (Fig.~\ref{fig:muller}). Any rounding error excites a second solution of the recurrence that tends to 100 and grows about seventeen-fold per step relative to the true one. In fixed-precision arithmetic with about a hundred digits or fewer the computed $x_{100}$ is therefore near 100, and only beyond roughly 120 digits does it settle near 6. With 300 digits the computed value is correct, but neither output gives any indication of its accuracy.

By contrast, iterating Eq.~\eqref{eq:muller} in ball arithmetic gives the same midpoints as the fixed-precision computation, each accompanied by a radius. At 16 and 50 digits of working precision a divisor ball eventually contains zero and the radius becomes infinite; the ball still contains the true value but guarantees no digit of it. Asked for 16 digits of $x_{100}$, Baller raises its working precision from 30 to 240 decimal digits and certifies $x_{100}=6.0000000160995649$; the same program in ordinary floating point returns 100.

Computing with more digits than seem necessary is the usual precaution, but the result carries no bound on how many of its digits are correct, and a single subtraction of nearly equal quantities can cancel thirty of them. Rerunning at higher precision to see whether the answer changes is impractical for a computation of millions of operations. In ball arithmetic the radius bounds the true error after each operation, so that the working precision can be raised exactly where a ball becomes too wide and the final enclosure can be used in a proof.

In particular, an interval-Newton step, the Krawczyk test~\cite{Krawczyk1969}, can prove that a given box (a product of intervals) contains exactly one zero of a system of equations. Applied to the stationarity or equilibrium conditions of a model, it yields a proof that an optimum or an equilibrium, first located numerically, exists and is unique within the box. The test requires a preconditioning matrix. A poorly chosen matrix may make it fail, but when it passes, the conclusion is proved whatever matrix was used. We use the test for the equilibria of a minimal model of atmospheric oxygen~\cite{GreatoxidationNote}. The same test applies to the critical point of the superpotential at a candidate string vacuum, whose existence depends on inequalities among periods known only numerically~\cite{StringvacNote}. However, enclosures can establish inequalities only: two disjoint balls prove that two numbers differ, whereas two overlapping balls of positive radius are consistent both with equal and with slightly different values of the two numbers. Equalities in our work come instead from exact rational arithmetic, or from an identity proposed by an integer-relation fit, Eq.~\eqref{eq:intrel}, and then either proved or accepted on passing the numerical test of Eq.~\eqref{eq:accept}.

\subsection{Correlated factors and Taylor forms\label{sec:eras}}

Naive ball arithmetic can fail by overestimation, returning an enclosure that is correct but far too wide to use. The common case in statistical work is a likelihood formed as a product of hundreds of per-observation factors that all depend on one shared parameter (a rate, say). Interval evaluation treats the factors as unrelated and adds the magnitudes of their first-order variations, which in the true product largely cancel. In one likelihood of this form the summed variations gave an enclosure of width $10^{22}$, although the likelihood's true sensitivity to the parameter was a few hundred.

The remedy for this overestimation is a classical device, the centered Taylor form~\cite{BerzMakino,Tucker2011}. It carries the dependence on the shared parameter as a truncated series of stated order with a rigorous bound on the remainder, and the cancellations then occur among series coefficients before any interval is widened. We packaged this construction as a routine, ERAS (for exact remainder-aware series), together with a set of test problems with known answers that any adaptation of it must reproduce. The routine was written when the failure first appeared, in a likelihood computation on linguistic data. It has since been reused for the arithmetic of the statistical applications of Section~\ref{sec:evidence}, among them the mixture-model and population-genetics computations~\cite{MixturesNote,MutationflowNote}.

\subsection{Audits of published numbers\label{sec:audits}}

A second use of exact and ball arithmetic is the audit of numbers already published (regulatory thresholds, quality ratings, reference tables). Such an audit again yields either proven digits or a refusal, and we report its result as one of four classes: the number in question can be reproduced from the published inputs; it is determined only under unstated conventions; it is undetermined by what was published; or it cannot be examined at all. When our recomputation disagrees with a published value, we state the criterion applied, the step of the calculation at which the discrepancy arises, and its size, in enough detail for a reader to reproduce the discrepancy from the same inputs. A routine written for the purpose, Recount, replays published arithmetic in exact rational numbers and thereby excludes our own rounding as a source of discrepancy.

Two examples from public health concern published numbers defined by a stated rule. The cut points of Medicare's star-rating systems are defined by the clustering criterion of Section~\ref{sec:complete}, and whether a published cut point coincides with the optimum of that criterion can be decided exactly~\cite{HealthstarsNote}. United States law limits ozone in outdoor air to 70 parts per billion. Whether a monitoring site attains the standard is judged by a three-year statistic of its readings (the design value), which under the regulation is truncated at its last digit after digits have already been discarded at earlier steps of the calculation. For a site near the limit, attainment can turn on the final digit alone. To determine how truncation rather than rounding affects attainment, we replay the whole chain of calculations in exact arithmetic from the public monitoring data~\cite{AirqualityNote}.

For another example, consider population genetics. There, the standard programs that infer the distribution of fitness effects from observed site-frequency spectra, among them polyDFE~\cite{polyDFE2017}, dadi~\cite{dadi2009} and fastDFE~\cite{fastDFE2024}, work in floating-point arithmetic. The exact spectra of Section~\ref{sec:areas} supply reference values against which the range of validity of such a program may be mapped~\cite{MutationflowNote}. The operating characteristics of a Bayesian clinical-trial design depend on Beta and Dirichlet tail integrals, which can likewise be computed with proven error bounds; Ref.~\cite{HealthtrialsNote} shows what exact arithmetic adds to the analysis of such trials.

%% file: s7_scope.tex
\section{Conclusions \label{sec:outlook}}

The thesis of this paper is that there is tremendous and largely underutilized potential in the breadth of knowledge of agentic AI. When this knowledge is mixed with the ability of the current generation of AI tools to write computer code and do technically challenging computations in mathematical physics, the combination is strikingly powerful. This viewpoint does not attempt to push out the deepest boundaries of human knowledge in one direction, but rather combines knowledge that already exists in different fields. That is, it allows us to fill in the convex hull of humanity's scientific corpus. This hull is enormous and much of it is unvisited: it contains every combination of a method from one field with a question from another, and in many pairings there is hardly anyone who knows both. The reduction algorithms of collider physics and the evidence integrals of phylogenetics are one example, interval arithmetic and population genetics another. A model that has read everything can bring everything together.

BootLoops is a first attempt at bringing together tools from distant areas to make it easier for scientists to make progress on long-standing problems, or to find new ones. BootLoops is meant to grow, as more methods are found and incorporated. Or other similar harnesses may be built elsewhere with other tools for other things. The hull is convex, but the possibilities are unbounded.

BootLoops is discussed in more detail at \url{https://bootloops.ai}. The code and forks of public programs are at \url{https://github.com/BootLoops-ai}.